\documentclass[12pt]{article}

\usepackage[margin=0.8in]{geometry}
\usepackage{setspace}
\usepackage{amsmath}
\usepackage{amssymb}
\usepackage{amsfonts}
\usepackage{amsthm}
\usepackage{bm}
\usepackage{dsfont}
\usepackage{mathrsfs}

\usepackage{graphicx}
\usepackage{booktabs}
\usepackage{array}
\usepackage{multirow}
\usepackage{longtable}
\usepackage{adjustbox}
\usepackage{rotating}
\usepackage{lscape}
\usepackage{caption}
\usepackage{subcaption}
\usepackage{placeins}

\usepackage{algorithm}
\usepackage[noend]{algpseudocode}

\usepackage{enumitem}
\usepackage{xcolor}

\usepackage[round]{natbib}
\usepackage[colorlinks=true,
            linkcolor=blue,
            citecolor=blue,
            urlcolor=blue]{hyperref}

\usepackage[switch, modulo]{lineno}

\makeatletter
\def\BState{\State\hskip-\ALG@thistlm}
\makeatother

\newtheorem{theorem}{Theorem}

\newtheorem{proposition}{Proposition}

\title{\bf Supervised Low-Rank Structure Discovery for Developmental Epigenetic Aging in Ultra-High-Dimensional DNA Methylation Data}

\author{
Priyam Das$^{1}$\thanks{Corresponding author. Email: \href{mailto:dasp4@vcu.edu}{dasp4@vcu.edu}}  \quad
Jiyeon Song$^{2}$  \quad
Lathika Mohanraj$^{3}$  \quad
Karolina A. Aberg$^{4}$\\[0.4em]
Yi Li$^{2}$  \quad
Subharup Guha$^{5}$ \\[0.6em]
{\normalsize $^{1}$Department of Biostatistics, Virginia Commonwealth University}\\
{\normalsize $^{2}$Department of Biostatistics, University of Michigan}\\
{\normalsize $^{3}$Department of Adult Health and Nursing Systems, Virginia Commonwealth University} \\
{\normalsize $^{4}$Center for Biomarker Research and Precision Medicine, Virginia Commonwealth University}\\
{\normalsize $^{5}$Department of Biomedical Data Science, Dartmouth College}
}

\date{}
       
\begin{document}
\maketitle
\begin{abstract}
Ultra-high-dimensional array-based CpG methylation studies require statistical frameworks that simultaneously provide supervised structure discovery, interpretability, scalable latent-dimension identification, and computational feasibility. We propose SOLAR (Supervised Orthogonal Low-rank Adaptive Regression), a supervised low-rank latent-factor framework for identifying   CpG-level methylation structure associated with residualized DNAm age. SOLAR combines orthogonal low-rank regression with a penalized maximum a posteriori formulation, dimension-adaptive BIC-type penalization, and a trans-dimensional simulated-annealing strategy for automatic latent-rank selection, together with theoretical guarantees including identifiability, fixed-rank recovery, and rank-selection consistency under suitable regularity conditions. The framework additionally incorporates computationally and memory-efficient optimization strategies demonstrating scalability up to \(p=10^7\), while analyses at \(p=10^6\) remain feasible on standard desktop computing environments. Simulation studies demonstrate stable rank recovery, competitive supervised signal recovery, and strong scalability across moderate-, high-, and ultra-high-dimensional regimes. Using longitudinal EPIC-array CpG methylation data from the GUSTO birth cohort, comprising \(n=1051\) methylation profiles collected across infancy and early childhood with approximately 860,000 assayed CpGs per sample, SOLAR identifies heterogeneous supervised methylation structure associated with residualized DNAm age beyond chronological age alone, together with biologically coherent CpG signatures and enrichment patterns.
\end{abstract}

\clearpage
\setstretch{1.7}
\section{Introduction}
\label{sec:intro}
DNA methylation (DNAm) has emerged as one of the most widely studied epigenomic mechanisms underlying human development, aging, and disease progression. By regulating gene expression without altering the underlying DNA sequence, DNAm plays a central role in cellular differentiation, developmental programming, immune regulation, neuro-development, and age-related physiological decline \citep{feinberg2018epigenetic,jones2012functions}. High-throughput methylation-array technologies now allow profiling of hundreds of thousands of CpG loci simultaneously \citep{pidsley2016critical}, enabling increasingly detailed investigation of biological aging processes across the life course. Among the most influential developments in this area are DNAm-based epigenetic clocks, which combine methylation measurements across selected CpG sites to estimate biological age  using defined algorithms \citep{horvath2013dna,hannum2013genome,levine2018epigenetic}. DNAm-based epigenetic age estimates and age-adjusted derivative measures have demonstrated strong associations with mortality, frailty, cardiometabolic disease, inflammation, cognitive decline, and other aging-related outcomes across multiple tissues and populations \citep{jylhava2017biological}.

Although epigenetic aging measures were originally developed primarily in adult cohorts, growing evidence suggests that early-life methylation dynamics may already contain meaningful signatures of developmental biological aging and long-term health heterogeneity \citep{felix2019dohad,simpkin2015longitudinal}. Early childhood represents a particularly important developmental window characterized by rapid physiological and epigenomic remodeling, where environmental exposures, prenatal conditions, stress, nutrition, inflammation, and socioeconomic factors may contribute to heterogeneous methylation trajectories and biological aging variation beyond chronological age alone \citep{felix2019dohad,crimmins2024generations}. Consequently, developmental epigenetic aging studies increasingly seek to identify structured methylation signatures associated with biological aging heterogeneity rather than merely estimating chronological age itself.

A major statistical challenge in such studies arises from the extreme dimensionality and dependence structure of modern methylation data. Contemporary Illumina methylation arrays routinely profile approximately 940K CpG sites per sample, while cohort sizes often remain in the hundreds or low thousands \citep{leroy2025longitudinal}. The resulting `large-\(p\), small-\(n\)' regime creates substantial computational and inferential difficulties for conventional regression and latent-factor methodologies. Moreover, methylation measurements exhibit strong local and global correlation structures induced by genomic proximity, shared regulatory mechanisms, and coordinated developmental processes \citep{zhou2017comprehensive}. These characteristics naturally motivate low-rank latent representations capable of recovering lower-dimensional methylation programs underlying observed biological heterogeneity.

Classical unsupervised dimension-reduction approaches such as principal component analysis (PCA) and principal component regression (PCR; \citealp[]{jolliffe2002principal}) recover dominant variance directions in high-dimensional data, but these directions need not align with biologically relevant aging variation. Outcome-informed methods, including partial least squares (PLS; \citealp[]{wold1984pls}) and supervised principal components (SPC; \citealp[]{bair2006prediction}), as well as broader supervised low-rank and reduced-rank regression formulations \citep{reinsel1998multivariate}, have been widely used in genomic and multivariate prediction settings. Related sparse factor, Bayesian latent-factor, matrix factorization, and supervised representation-learning approaches have also been developed for moderate-to-high-dimensional omics studies \citep{bhattacharya2011sparse,carvalho2008high,fan2014challenges}. However, several challenges remain insufficiently addressed for developmental epigenetic aging studies. First, many supervised dimension-reduction methods are primarily prediction-oriented, including PLS, SPC, and supervised PCA-type formulations \citep{wold1984pls,bair2006prediction,barshan2011supervised}, whereas the present scientific goal requires recovery of interpretable methylation programs associated with biological-aging heterogeneity. Second, latent-dimension selection remains challenging in ultra-high-dimensional supervised factor models. Existing reduced-rank regression approaches typically select rank through cross-validation, information criteria, stability analysis, or penalized singular-value formulations \citep{bunea2011optimal,chen2013reduced}. Stability-based methods such as StARS-RRR provide important advances for low-to-moderate dimensional reduced-rank regression \citep{wen2023stability}, but are not designed for supervised latent-structure discovery at methylation-array scale. Third, computational scalability remains a central obstacle: recent sparse reduced-rank regression methods for omics integration provide useful interpretability but have typically been evaluated on transcriptomic datasets with at most a few thousand genes rather than genome-wide methylation settings \citep{kobak2021sparse}. Finally, existing supervised latent-factor approaches are often not designed to produce CpG-level loading structures for downstream biological characterization while remaining computationally feasible at EPIC-array scale.

Motivated by these challenges, we develop SOLAR (Supervised Orthogonal Low-rank Adaptive Regression), a supervised latent-factor framework for identifying biologically interpretable methylation structure associated with residualized DNAm age during early childhood. SOLAR combines orthogonal low-rank factor regression with a penalized maximum a posteriori formulation, dimension-adaptive BIC-type penalization, and a trans-dimensional simulated-annealing strategy for adaptive latent-rank selection, while incorporating computationally and memory-efficient optimization strategies enabling scalability in ultra-high-dimensional settings up to \(p=10^7\). Unlike purely unsupervised low-rank approaches, SOLAR explicitly incorporates biological aging outcomes during latent-factor estimation to recover supervised methylation programs, CpG-level loading organization, and biologically coherent downstream enrichment structure, together with theoretical guarantees for identifiability, fixed-rank recovery, and rank-selection consistency. The framework is motivated by methylation data from the longitudinal GUSTO birth cohort \citep{leroy2025longitudinal}, comprising (n=1051) EPIC-array CpG methylation profiles collected across infancy and early childhood, with approximately 860,000 assayed CpG sites per profile. Although the cohort includes repeated measurements over developmental time, the present analysis uses the available methylation profiles to identify supervised CpG-level structure associated with residualized DNAm age, rather than explicitly modeling subject-specific longitudinal methylation trajectories. Specifically, we derive residualized DNAm age measures from established epigenetic clocks, including Horvath, Hannum, and Levine/DNAm PhenoAge \citep{horvath2013dna,hannum2013genome,levine2018epigenetic}, and investigate whether lower-dimensional supervised methylation structure explains variation in DNAm age beyond chronological age during early childhood.

The remainder of the paper is organized as follows. Section~\ref{sec:realdata} introduces the developmental epigenetic aging problem and motivating exploratory analyses. Section~\ref{sec:solar} develops the SOLAR framework, while Section~\ref{sec:theory} summarizes the main theoretical results. Section~\ref{sec:sim_study} presents simulation studies across multiple dimensional regimes, and Section~\ref{sec:realdata_results} presents the developmental epigenetic aging analysis. Section~\ref{sec:discussion} concludes with discussion and future directions.

\section{Developmental Epigenetic Aging in Early Childhood}\label{sec:realdata}
DNAm-based epigenetic clocks combine methylation levels across selected CpG loci to estimate biological age and have been associated with mortality, frailty, cardiometabolic disease, cognitive decline, inflammation, and other aging-related outcomes across diverse populations and tissues \citep{horvath2013dna,hannum2013genome,levine2018epigenetic,crimmins2024generations}. Among the most widely studied clocks are the multi-tissue Horvath clock, the blood-based Hannum clock, and the phenotypic aging--oriented Levine/DNAm PhenoAge clock, each constructed using distinct training objectives and CpG subsets. Although these clocks were primarily developed and validated in adult populations, increasing evidence suggests that early-life methylation dynamics may contain meaningful signatures of developmental biological aging and long-term health trajectories \citep{han2024lagged,moqri2024prc2,clark2024impact,copeland2022early,jansen2021integrative,han2018epigenetic}.

Early childhood is a key developmental window in which rapid physiological, immunological, metabolic, and neurodevelopmental changes occur alongside substantial epigenomic remodeling. Environmental exposures, prenatal conditions, nutrition, stress, inflammation, and developmental processes may contribute to heterogeneous methylation trajectories during childhood \citep{felix2019dohad,simpkin2015longitudinal}. Consequently, chronological age alone may inadequately characterize developmental biological heterogeneity, motivating supervised low-dimensional representations that identify structured methylation variation associated with developmental epigenetic aging.

We analyze publicly available longitudinal methylation data from GEO accession GSE254135 \citep{leroy2025longitudinal}, a pediatric birth cohort with DNA methylation profiled during early childhood using Illumina methylation arrays. The original study developed probabilistic longitudinal prediction models for forecasting future methylation trajectories across childhood; here, we instead focus on developmental biological aging heterogeneity and supervised latent methylation structure. Methylation measurements were available across four developmental age groups, approximately 3, 9, 48, and 72 months. Preprocessing and quality control followed the original data source \citep{leroy2025longitudinal}; the retained profiles span infancy through early childhood in the longitudinal GUSTO birth cohort.

\begin{figure}[!t]
\centering
\includegraphics[width=0.9\textwidth]{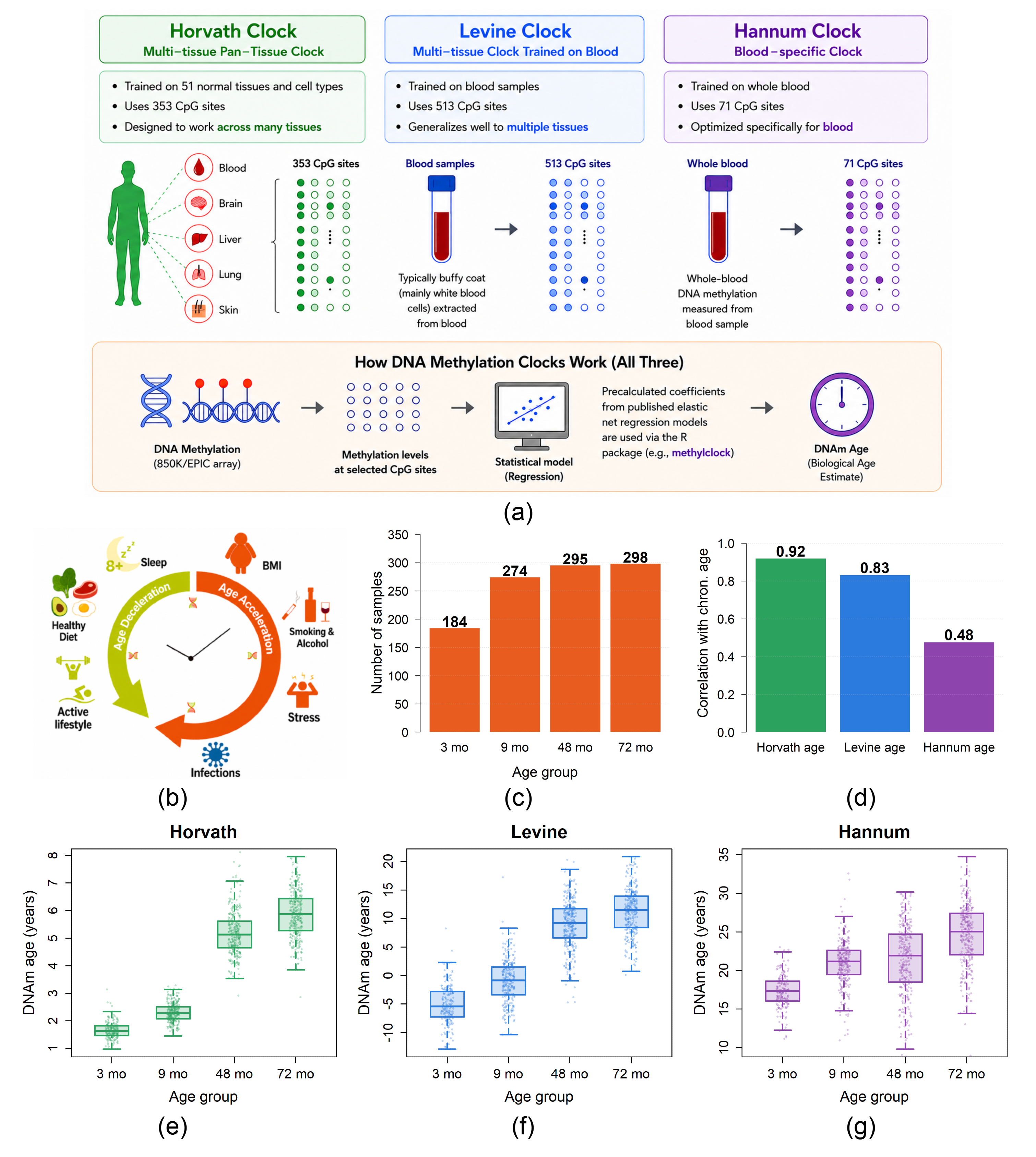}
\caption{
Motivational overview and exploratory characterization of developmental epigenetic aging in the GSE254135 early-childhood methylation cohort. 
(a) Overview of the three DNAm clocks considered in this study: the multi-tissue Horvath clock (353 CpGs), Levine/DNAm PhenoAge (513 CpGs), and the blood-specific Hannum clock (71 CpGs). 
(b) Conceptual illustration of biological age acceleration and deceleration relative to chronological aging. 
(c) Sample-size distribution across the four developmental age groups (3, 9, 48, and 72 months). 
(d) Correlations between chronological age and the three DNAm aging measures, illustrating heterogeneous developmental calibration across clocks. 
(e)--(g) Age-group–specific distributions of Horvath, Levine/DNAm PhenoAge, and Hannum DNAm age estimates, respectively.
}
\label{fig:motivation}
\end{figure}

To characterize developmental epigenetic aging, we compute three widely used DNAm aging measures: Horvath DNAm age, Hannum DNAm age, and Levine/DNAm PhenoAge. Figure~\ref{fig:motivation}(a) summarizes the corresponding epigenetic clocks and their differing biological objectives, tissue specificity, and CpG compositions. The Horvath clock was developed across multiple tissues using 353 CpGs \citep{horvath2013dna}, the Hannum clock was trained on whole blood using 71 CpGs \citep{hannum2013genome}, and Levine/DNAm PhenoAge incorporates mortality- and physiology-related phenotypic aging information using 513 CpGs \citep{levine2018epigenetic}. All three clocks estimate DNAm age through previously trained regression models combining methylation measurements across selected CpG sites.

Figure~\ref{fig:motivation}(b) illustrates the conceptual distinction between chronological age and biological age acceleration, where heterogeneous environmental, behavioral, physiological, and developmental influences may induce substantial variability in epigenetic aging trajectories beyond age alone. Figures~\ref{fig:motivation}(c)--(g) summarize exploratory characteristics of the cohort and the resulting DNAm age measures. Figure~\ref{fig:motivation}(c) shows the sample distribution across developmental age groups. Figure~\ref{fig:motivation}(d) displays correlations between chronological age and the three DNAm aging measures, revealing substantial heterogeneity in developmental calibration across clocks. Consistent with prior observations regarding pediatric applications of adult-trained clocks, Horvath DNAm age demonstrated the strongest developmental coherence with chronological age (\(r=0.92\)), followed by Levine/DNAm PhenoAge (\(r=0.83\)), whereas Hannum DNAm age showed comparatively weaker calibration in this early-life setting (\(r=0.48\)). Figures~\ref{fig:motivation}(e)--(g) further illustrate age-group–specific DNAm age distributions for each clock. The observed dispersion patterns suggest that developmental methylation heterogeneity cannot be fully explained by chronological age alone and may instead involve lower-dimensional latent biological structure.

Because the primary scientific interest lies in methylation-based biological age variation beyond chronological age, subsequent analyses use residualized DNAm age measures, which are widely used in the biological aging literature as indicators of aging-related deviation beyond chronological age alone \citep{horvath2013dna,levine2018epigenetic}. Specifically, for each clock, chronological age was regressed out from the corresponding DNAm age estimate, and the resulting residuals were used as supervised outcomes. We interpret these residualized DNAm age measures cautiously, since established epigenetic clocks may be imperfectly calibrated in early-childhood samples; therefore, the exploratory analyses above are used to assess clock behavior in this cohort before selecting the primary outcome. These observations motivate the central scientific questions of this work:
\begin{enumerate}
\item Do early-life DNA methylation profiles reveal heterogeneous developmental patterns of methylation-based biological age variation beyond chronological age alone?
\item To what extent can low-dimensional supervised methylation structure explain variation in residualized DNAm age during early childhood?
\item Which methylation signatures contribute most strongly to residualized methylation-based biological age variation, and are these signatures consistent across distinct epigenetic aging measures?
\end{enumerate}

\noindent To address these questions, we develop a supervised orthogonal low-rank framework, termed SOLAR, for extracting biologically interpretable latent methylation structure associated with developmental epigenetic aging variation. Unlike purely unsupervised dimension-reduction approaches, the proposed framework explicitly incorporates developmental biological aging outcomes during latent factor estimation, thereby encouraging recovery of methylation structure most relevant to epigenetic aging heterogeneity. Because methylation measurements are ultra-high dimensional and highly correlated, supervised low-dimensional representations may provide substantially improved interpretability and biological coherence relative to black-box predictive models, such as deep-learning approaches, alone. Moreover, the dimensional scale of modern methylation arrays poses substantial computational and statistical challenges for conventional supervised latent-factor methods, particularly when stable low-rank estimation and interpretable factor recovery are desired simultaneously.

\section{Penalized MAP Estimation with Trans-Dimensional Stiefel Optimization}\label{sec:solar}
We develop a computationally efficient penalized maximum a posteriori (MAP) estimator for the supervised low-rank regression model, where \(X\in\mathbb R^{n\times p}\) denotes the array-based CpG methylation profile matrix and \(w\in\mathbb R^n\) denotes the supervised outcome, such as residualized DNAm age. Rather than performing full posterior sampling, we exploit the conjugate Gaussian structure to obtain a structured penalized likelihood formulation with automatic rank regularization. Consider the supervised factor representation
\begin{align*}
X = H D V^\top + U, \; w = H\beta + \varepsilon,
\end{align*}

\noindent where $H \in \mathbb{R}^{n\times q}$ and $V \in \mathbb{R}^{p\times q}$ lie on the Stiefel manifolds $H^\top H = I_q$, $V^\top V = I_q$, $D=\mathrm{diag}(d_1,\dots,d_q)$, and $\beta \in \mathbb{R}^q$.
We assume Gaussian noise,
\[ U_{ij}\stackrel{\text{iid}}{\sim}\mathcal{N}(0,\sigma^2),
\qquad
\varepsilon_i\stackrel{\text{iid}}{\sim}\mathcal{N}(0,\tau^2).
\]

\noindent To regularize the latent singular values and supervised regression coefficients, we impose independent Gaussian priors
\[
d \sim \mathcal{N}(0,\rho^2 I_q),
\qquad
\beta \sim \mathcal{N}(0,g^2 I_q).
\]

\noindent Although the prior on \(d\) is written on the unconstrained scale, the fitted diagonal entries of \(D\) are represented canonically as nonnegative singular-value magnitudes through the SVD-based basis rotation described below. The orthogonality constraints are enforced by assigning $(H,V)$ uniform (Haar) priors on their respective Stiefel manifolds. 
These priors introduce no additional penalty beyond the manifold constraints and therefore act as noninformative regularization on the orientation of the latent subspace.
Under these assumptions, the log-posterior for fixed rank $q$, up to additive constants, is
\begin{align}
\ell(H,V,D,\beta;q)
=
-\frac{1}{2\sigma^2}\|X - H D V^\top\|_F^2
-\frac{1}{2\tau^2}\|w - H\beta\|_2^2 
-\frac{1}{2\rho^2}\|d\|_2^2
-\frac{1}{2g^2}\|\beta\|_2^2.
\label{eq:logpost}
\end{align}

\noindent \textbf{Penalized Objective and Rank Regularization:}
The log-posterior \eqref{eq:logpost} defines a regularized likelihood for fixed rank $q$. 
To enable adaptive rank selection, we augment it with a BIC-type structural penalty and define the penalized objective
\begin{align}
J(H,V,D,\beta,q)
=
\ell(H,V,D,\beta)
-
\mathrm{pen}(q),
\label{eq:J_def}
\end{align}
where
\begin{align}
\mathrm{pen}(q)
=
\kappa \cdot \frac{1}{2}\log(np)\,\mathrm{df}(q),
\qquad
\mathrm{df}(q)= q(n+p-2q)+2q.
\label{eq:penalty}
\end{align}

The factor $\log(np)$ follows the classical BIC principle, since under the Gaussian model in Section~2 the log-likelihood is a sum over $np$ observations, yielding effective sample size $np$. The quantity $\mathrm{df}(q)$ represents the intrinsic dimension of the parameter space under orthogonality constraints. Specifically, $H$ and $V$ lie on the Stiefel manifolds $\mathbb{V}_q(\mathbb{R}^n)$ and $\mathbb{V}_q(\mathbb{R}^p)$, whose intrinsic dimensions are $nq - \frac{q(q+1)}{2}$ and $pq - \frac{q(q+1)}{2}$, respectively. Augmenting these by the $q$ diagonal loading parameters in $D$ and the $q$ supervised regression coefficients in $\beta$ yields $\mathrm{df}(q)$ as in \eqref{eq:penalty}. Thus, the SOLAR estimator is defined as the maximizer of \eqref{eq:J_def} over
$H \in \mathbb{V}_q(\mathbb{R}^n), \quad
V \in \mathbb{V}_q(\mathbb{R}^p), \quad
D \in \mathbb{R}^{q \times q} \text{ diagonal}, \quad
\beta \in \mathbb{R}^q,
$
with $1 \le q \le q_{\max}$. Under a direct BIC analogue one would take $\kappa=1$. However, in ultra-high-dimensional settings, fixed-penalty calibrations may become unstable due to the dependence of noise singular values on the ambient dimensionality. To improve finite-sample stability, we therefore allow the strength of the rank penalty to vary with $(n,p)$. Classical information criteria are typically calibrated for fixed-$p$ asymptotics, whereas in modern low-rank problems the scale of noise singular values depends strongly on the ambient dimensionality \citep{johnstone2001distribution}. Motivated by the corresponding high-dimensional spectral edge, we use the dimension-adaptive scaling
\[
\kappa(n,p)
=
c_{\kappa}
\left(
\frac{(\sqrt n+\sqrt p)^2}{n+p}
\right)^{\zeta},
\]

\noindent where $c_{\kappa}>0$ controls the overall regularization level and $\zeta>0$ governs the degree of dimensional adaptation. The term $(\sqrt n+\sqrt p)^2$ reflects the first-order scale of noise singular values in high-dimensional random matrices \citep{johnstone2001distribution}, while normalization by $(n+p)$ adjusts for the increasing complexity of the latent rank structure. Consequently, the penalty becomes relatively stronger in lower-dimensional finite-sample settings while remaining stable when $p\gg n$, thereby improving empirical rank recovery across a broad range of dimensional regimes. In our experiments, we use the fixed values $c_{\kappa}=0.1$ and $\zeta=2/3$. Across a broad range of moderate- to ultra-high-dimensional simulation settings and sample sizes, this calibration provided stable empirical rank recovery and competitive finite-sample behavior. While a formal theoretical characterization of the optimal scaling remains beyond the scope of the current work, further analysis of such dimension-adaptive penalty structures may be of independent methodological interest.
\subsection{Optimization Strategy}
Maximization of the penalized objective $J(H,V,D,\beta,q)$ proceeds in two layers. 
We first consider the fixed-rank problem for a given $q$, where the objective is optimized over the constrained parameter space $(H, V, D, \beta) \in \mathbb{V}_q(\mathbb{R}^n)
\times
\mathbb{V}_q(\mathbb{R}^p)
\times
\mathbb{R}^q
\times
\mathbb{R}^q.$
Within this fixed-dimensional setting, we employ block-wise alternating updates, each admitting a closed-form solution. 
Rank updates are subsequently incorporated through a trans-dimensional search described below. To improve scalability, objective evaluations are implemented using trace identities that avoid explicit construction of the fitted matrix \(HDV^\top\).

\subsubsection{Fixed-Rank Alternating MAP Updates}
For fixed $q$, maximization of $J$ is carried out by iteratively updating $(V,H,\beta,D)$ while respecting the orthogonality constraints.
For high- and ultra-high-dimensional settings with \(p\gg n\), initialization and rank proposals are computed using a Gram-matrix eigen-decomposition \citep{golub2013matrix} of \(XX^\top\) rather than a direct singular value decomposition of \(X\). The corresponding right singular vectors are recovered through back-projection, substantially reducing memory and computational burden in ultra-high-dimensional regimes. For fixed rank \(q\), initialization is based on the leading rank-\(q\) spectral approximation of \(X\). Let \(H^{(0)}\) denote the leading \(q\) eigenvectors of \(XX^\top\), and let \(s_1,\ldots,s_q\) be the corresponding singular values. We set \(V^{(0)} = X^\top H^{(0)}\operatorname{diag}(s_1^{-1},\ldots,s_q^{-1})\), \(D^{(0)} = \{\sigma^{-2}/(\sigma^{-2}+\rho^{-2})\}\operatorname{diag}(s_1,\ldots,s_q)\), and \(\beta^{(0)} = \{\tau^{-2}/(\tau^{-2}+g^{-2})\}(H^{(0)})^\top w\). These initial values are then refined by the block-wise updates below.

\paragraph{Update of $V$.}
Conditioning on $(H,D)$, the objective reduces to
\[
\max_{V^\top V=I_q}
\operatorname{tr}(V^\top X^\top H D),
\]

\noindent an orthogonal Procrustes problem. Let $X^\top H D = U_1 \Sigma_1 V_1^\top$ be its thin SVD. 
The maximizer is $V \leftarrow U_1 V_1^\top$.
\paragraph{Update of $H$.}
Conditioning on $(V,D,\beta)$, maximization becomes
\[
\max_{H^\top H=I_q}
\operatorname{tr}\!\left(
H^\top
\left(
\frac{X V D}{\sigma^2}
+
\frac{w\beta^\top}{\tau^2}
\right)
\right).
\]

\noindent Let
$\frac{X V D}{\sigma^2}
+
\frac{w\beta^\top}{\tau^2}
=
U_2 \Sigma_2 V_2^\top$
be its thin SVD. 
The maximizer is $H \leftarrow U_2 V_2^\top$.
\paragraph{Update of $\beta$.}
Given $H$, the Gaussian prior yields the shrinkage estimator
\[
\beta
=
\frac{\tau^{-2}}{\tau^{-2}+g^{-2}}
H^\top w.
\]

\paragraph{Update of $D$ (basis rotation).}
Given $(H,V)$, define $R = H^\top X V \in \mathbb{R}^{q\times q}$,
and compute its singular value decomposition $R = U_3 \Sigma V_3^\top$. We perform a change of orthonormal basis within the latent space via $H \leftarrow H U_3$, $V \leftarrow V V_3$,
and update
\[
D \leftarrow
\frac{\sigma^{-2}}{\sigma^{-2}+\rho^{-2}}
\operatorname{diag}(\Sigma).
\]

\noindent This transformation preserves the column spaces of $H$ and $V$ while enforcing a diagonal loading matrix.
\paragraph{Post-update rotational stabilization.} Before imposing a canonical orientation, the latent representation is non-identifiable up to orthogonal reparameterizations of the latent subspace:
\[
(H,V,\beta) \mapsto (H Q, V Q, Q^\top \beta),
\qquad Q \in \mathcal{O}(q).
\]
Consequently, the penalized objective $J$ depends only on the column spaces of $H$ and $V$ and the induced linear predictor $H\beta$, not on the particular orthonormal basis used to represent the latent subspace. 

Although this invariance does not affect fitted values or the value of $J$, it introduces equivalent maximizers that may produce arbitrary rotations of the latent basis across iterations, complicating convergence diagnostics and interpretation of the supervised direction. To remove this ambiguity, we anchor the latent basis to the supervised signal after each alternating cycle. Let
\[
u = H^\top w,
\qquad
a = \frac{u}{\|u\|}.
\]

\noindent We construct a Householder matrix $R_w$ satisfying $R_w e_1 = a$, and apply the joint rotation
\[
H \leftarrow H R_w^\top,
\qquad
V \leftarrow V R_w^\top,
\qquad
\beta \leftarrow R_w \beta.
\]

\noindent This transformation leaves $J$ and all fitted values unchanged while stabilizing the latent orientation relative to the supervised signal. Combined with the ordering induced by \(D\), it removes the remaining sign and ordering ambiguity and yields a deterministic representative of the fitted factorization.

\subsubsection{Trans-Dimensional Annealed Rank Updates}
After each alternating optimization cycle at fixed rank $q$, we allow the dimension to change through a stochastic rank-update mechanism. This step enables automatic rank selection while preserving the deterministic structure of the fixed-rank updates. Let $J_t$ denote the current penalized objective value at iteration $t$. We consider local rank proposals of the form
\[
q^\star = q + \delta, 
\qquad \delta \in \{-1,+1\},
\]

\noindent
subject to $q_{\min} \le q^\star \le q_{\max}$. The proposal direction $\delta$ is chosen with equal probability. For a proposed rank \(q^\star\), the algorithm reuses a precomputed spectral initialization based on the leading singular vectors of \(X\), together with the corresponding closed-form MAP update for \(\beta^\star\). The associated penalized objective value is denoted $J^\star$. Let $t=1,2,\dots$ index the outer iterations of the algorithm, 
each consisting of one fixed-rank alternating cycle followed by a possible rank proposal. 
Acceptance of the proposed rank is governed by a simulated annealing scheme. 
We define the temperature schedule as
\[
T_t = T_0\, t^{-\gamma}, 
\qquad \gamma \in (0,1),
\]

\noindent where $T_0>0$ is an initial temperature. The proposed configuration is accepted with probability
\[
\alpha_t
=
\min\left\{
1,
\exp\!\left(
\frac{J^\star - J_t}{T_t}
\right)
\right\}.
\]

\noindent Hence, improvements in the penalized objective are always accepted, while inferior moves may be accepted with probability decreasing in both the objective gap and the temperature. 
This annealed acceptance mechanism encourages exploration of nearby ranks during early iterations and gradually concentrates the search near the maximizer of $J$ as $t$ increases. Throughout the procedure, we record the configuration attaining the largest penalized objective value encountered so far, which is reported as the final estimator.

The trans-dimensional stage is terminated when a stability criterion is satisfied. Specifically, termination requires (i) stabilization of the selected rank over a sliding window of iterations, (ii) negligible relative improvement of the best penalized objective within that window, and (iii) a vanishing empirical acceptance rate of rank proposals. This ensures that exploration has ceased and the procedure has concentrated near a stable maximizer. The precise stopping criteria and implementation details are explicitly provided in the Supplementary Section A.

\subsubsection{Final Estimator and Evaluation Alignment}
The final estimator is the parameter configuration achieving the maximal penalized objective during the annealing trajectory. For simulation evaluation only, we perform a post-hoc Procrustes alignment between estimated and true latent subspaces to compute subspace and coefficient discrepancies under rotational invariance. 
This alignment is not used during optimization and does not affect fitted values or predictive performance.
\section{Theoretical Properties}\label{sec:theory}
We briefly summarize the main theoretical properties of the proposed framework. Throughout, we assume the supervised factor model described in Section~\ref{sec:solar}, where
\[
X = H_0D_0V_0^\top + E,
\qquad
w = H_0\beta_0 + \varepsilon,
\]

\noindent with Gaussian noise and fixed latent rank \(q_0\). Theoretical development is carried out under standard regularity conditions for low-rank latent factor models: (A1) the nonzero diagonal entries of \(D_0\) are distinct, positive, and spectrally separated from the noise bulk; (A2) the supervised component is nondegenerate and identifiable under the Stiefel normalization; (A3) the ambient dimension may grow with \(n\), allowing \(p\ge n\), provided \(\log p=o(n)\), while \(q_0\) remains fixed; (A4) the dimension-adaptive penalty multiplier remains uniformly bounded away from zero and infinity; (A5) the candidate rank space is fixed and contains \(q_0\); and (A6) the weakest latent factor satisfies a signal-detectability condition sufficient for consistent rank selection under the BIC-type penalty. Complete assumptions, technical justification, and proofs are provided in the Supplementary Section B. The first result formalizes the identifiability structure of the supervised factor representation.
\begin{proposition}[Identifiability up to signed permutations]
\label{prop:identifiability}
Under assumptions (A1)--(A5), the supervised factor representation is identifiable up to simultaneous signed permutations of the latent factors. Moreover, the canonical orientation step introduced in Section~\ref{sec:solar} removes the remaining sign and ordering ambiguity in a deterministic way without altering the fitted matrix \(HDV^\top\), the fitted supervised signal \(H\beta\), or the value of the objective function.
\end{proposition}

\noindent Proposition~\ref{prop:identifiability} clarifies that the inferential objects targeted by SOLAR are the latent subspaces, the low-rank signal, and the supervised signal, all of which are invariant to arbitrary labeling and sign conventions for the latent factors. The next theorem establishes recovery of the latent structure when the true rank is known.

\begin{theorem}[Fixed-rank recovery]
\label{thm:fixed_rank}
Under assumptions (A1)--(A5), suppose the true latent rank \(q_0\) is known. Let \((\hat H,\hat V,\hat D,\hat\beta)\) denote any maximizer of the fixed-rank objective over the constrained parameter space, represented under the canonical orientation in Proposition~\ref{prop:identifiability}. Then, as \(n\to\infty\) with \(p=p_n\) allowed to grow according to (A3),
\[
\hat H\hat H^\top \xrightarrow{p} H_0H_0^\top,\quad
\hat V\hat V^\top \xrightarrow{p} V_0V_0^\top,\quad
\frac{\|\hat D-D_0\|_F}{\|D_0\|_F} \xrightarrow{p}0,\quad
\|\hat\beta-\beta_0\|_2\xrightarrow{p}0.
\]

\noindent Consequently, the low-rank signal \(HDV^\top\) is consistently recovered in relative Frobenius error, and the supervised signal \(H\beta\) is consistently recovered.
\end{theorem}

Theorem~\ref{thm:fixed_rank} shows that, when the latent dimension is correctly specified, the supervised factor formulation consistently recovers the sample-level latent structure, feature-level loading structure, and supervised component. The final result establishes consistency of the adaptive rank-selection procedure induced by the proposed penalized objective.

\begin{theorem}[Rank-selection consistency]
\label{thm:rank_consistency}
Under assumptions (A1)--(A6), let \((\hat H,\hat V,\hat D,\hat\beta,\hat q)\) maximize
\[
J(H,V,D,\beta,q)
=
\ell(H,V,D,\beta)
-
\kappa(n,p)\frac12\log(np)\,\mathrm{df}(q),
\]

\noindent over \(1\le q\le q_{\max}\). Then
\[
\Pr(\hat q=q_0)\to1.
\]
\end{theorem}

\noindent Theorem~\ref{thm:rank_consistency} provides theoretical support for the proposed dimension-adaptive penalization strategy by showing that, under the stated detectability conditions, the procedure asymptotically selects the true latent rank while avoiding systematic overfitting and underfitting.
\section{Simulation Study}\label{sec:sim_study}
We evaluated SOLAR across a range of moderate-, high-, and ultra-high-dimensional settings designed to assess latent rank recovery, supervised signal recovery, predictive performance, and computational scalability. Competing methods included principal component regression (PCR), partial least squares (PLS; \citealp[]{wold1984pls}), and supervised principal components (SPC; \citealp[]{bair2006prediction}). All results were averaged over 10 independent replications. Full details of the data-generation mechanism, implementation procedures, and explicit definitions of the evaluation metrics are provided in the Supplementary Section A; here we briefly summarize the overall simulation design. Data were generated from the supervised low-rank model
\[
X = H_{\mathrm{true}} D_{\mathrm{true}} V_{\mathrm{true}}^\top + E,
\qquad
w = H_{\mathrm{true}}\beta_{\mathrm{true}} + \varepsilon,
\]

\noindent where $H_{\mathrm{true}}$ and $V_{\mathrm{true}}$ were orthonormal latent factor matrices generated from Gaussian random matrices via QR decomposition, $E$ contained independent Gaussian noise, and $\beta_{\mathrm{true}}$ controlled the supervised latent signal. Signal strengths were controlled using matrix-level and supervised signal-to-noise ratios corresponding to three representative regimes: high-$X$/low-$w$ signal $(\mathrm{SNR}_X,\mathrm{SNR}_w)=(0.20,0.05)$, low-$X$/high-$w$ signal $(0.01,0.50)$, and a moderate signal setting $(0.05,0.20)$. Latent ranks $q\in\{3,5\}$ were considered throughout.

All simulations were implemented in MATLAB R2024a on AMD EPYC high-performance computing nodes. Moderate-, high-, and \(p=10^6\) ultra-high-dimensional experiments used up to 256 GB RAM, where individual model fits were performed using single-core execution while parallel resources were used to evaluate multiple simulation replications and tuning scenarios simultaneously. The \(p=10^7\) ultra-high-dimensional experiments were conducted on a separate 494 GB RAM node using 40-thread multi-threaded execution. Moderate- and high-dimensional experiments considered
$(n,p)\in\{(500,1000),$ $(500,10000),$
$(1000,10000),$ $(1000,100000)\}$.
Representative results for the settings $(500,1000)$ and $(1000,100000)$ are reported in the main paper, while results for the other two intermediate-dimensional settings are provided in the Supplementary Table S1. For these settings, each replication used an 80/20 train-test split. All methods were trained using the training set, while out-of-sample prediction performance was evaluated on held-out test data. PCR used principal components estimated from centered training predictors, PLS used MATLAB's \texttt{plsregress} implementation, and SPC followed the supervised screening framework of \citet{bair2006prediction}. For PCR, PLS, and SPC, latent dimension was selected over $q\in\{1,\ldots,10\}$ using training-based cross-validation; SPC additionally selected the screening proportion over $\{1\%,2.5\%,5\%,10\%,20\%\}$ of predictors. SOLAR used the same candidate rank range together with the adaptive penalized objective in Section~\ref{sec:solar}, enabling automatic rank selection without cross-validation. Performance was evaluated using latent rank recovery, $H$-projector error, supervised signal RMSE and correlation, out-of-sample RMSE for prediction of $w$, and runtime.

\begin{table}[!t]
\centering
\renewcommand{\arraystretch}{1.2}
\resizebox{0.9\columnwidth}{!}{%
\begin{tabular}{ccccccccccc}
\hline
$(n,p)$ & $q$ & SNR & Method & $\hat{q}$ & \begin{tabular}[c]{@{}c@{}}Rank recovery\\ prop.\end{tabular} & \begin{tabular}[c]{@{}c@{}}$H$-proj. \\ error\end{tabular} & \begin{tabular}[c]{@{}c@{}}Sup. signal\\ RMSE\end{tabular} & \begin{tabular}[c]{@{}c@{}}Sup. signal \\ corr.\end{tabular} & \begin{tabular}[c]{@{}c@{}}Test RMSE\\ ($w$)\end{tabular} & Time (s) \\ \hline
\multicolumn{1}{c|}{\multirow{24}{*}{$(500, 10^3)$}} & \multicolumn{1}{c|}{\multirow{12}{*}{3}} & \multicolumn{1}{c|}{\multirow{4}{*}{\begin{tabular}[c]{@{}c@{}}SNR\_X = 0.20\\ SNR\_w = 0.05\end{tabular}}} & PCR & 2.20 (0.33) & 0.6 (0.16) & \textbf{0.383 (0.036)} & 0.183 (0.027) & 0.786 (0.070) & 1.434 (0.030) & 0.10(0.00) \\
\multicolumn{1}{c|}{} & \multicolumn{1}{c|}{} & \multicolumn{1}{c|}{} & PLS & 1.00 (0.00) & 0.0 (0.00) & 0.987 (0.055) & 0.277 (0.011) & 0.930 (0.012) & 1.446 (0.032) & 0.2 (0.00) \\
\multicolumn{1}{c|}{} & \multicolumn{1}{c|}{} & \multicolumn{1}{c|}{} & SPC & 4.20 (0.49) & 0.2 (0.13) & 0.845 (0.054) & 0.290 (0.019) & 0.863 (0.019) & 1.442 (0.031) & 1.0 (0.02) \\
\multicolumn{1}{c|}{} & \multicolumn{1}{c|}{} & \multicolumn{1}{c|}{} & SOLAR & 3.00 (0.00) & \textbf{1.0 (0.00)} & 0.453 (0.009) & \textbf{0.102 (0.009)} & \textbf{0.959 (0.007)} & \textbf{1.425 (0.029)} & 1.1 (0.01) \\ \cline{3-11} 
\multicolumn{1}{c|}{} & \multicolumn{1}{c|}{} & \multicolumn{1}{c|}{\multirow{4}{*}{\begin{tabular}[c]{@{}c@{}}SNR\_X = 0.01\\ SNR\_w = 0.50\end{tabular}}} & PCR & 2.90 (0.10) & \textbf{0.9 (0.10)} & 1.467 (0.019) & 0.220 (0.004) & 0.727 (0.011) & 0.505 (0.013) & 0.1 (0.01) \\
\multicolumn{1}{c|}{} & \multicolumn{1}{c|}{} & \multicolumn{1}{c|}{} & PLS & 1.00 (0.00) & 0.0 (0.00) & \textbf{1.131 (0.014)} & 0.294 (0.006) & \textbf{0.768 (0.006)} & 0.510 (0.011) & 0.2 (0.01) \\
\multicolumn{1}{c|}{} & \multicolumn{1}{c|}{} & \multicolumn{1}{c|}{} & SPC & 4.00 (0.92) & 0.2 (0.13) & 1.729 (0.115) & 0.274 (0.004) & 0.715 (0.005) & 0.517 (0.013) & 1.1 (0.01) \\
\multicolumn{1}{c|}{} & \multicolumn{1}{c|}{} & \multicolumn{1}{c|}{} & SOLAR & 2.70 (0.15) & 0.7 (0.15) & 1.336 (0.049) & \textbf{0.216 (0.005)} & 0.747 (0.015) & \textbf{0.503 (0.014)} & 0.9 (0.04) \\ \cline{3-11} 
\multicolumn{1}{c|}{} & \multicolumn{1}{c|}{} & \multicolumn{1}{c|}{\multirow{4}{*}{\begin{tabular}[c]{@{}c@{}}SNR\_X = 0.05\\ SNR\_w = 0.20\end{tabular}}} & PCR & 3.00 (0.00) & \textbf{1.0 (0.00)} & \textbf{0.672 (0.006)} & \textbf{0.100 (0.004)} & \textbf{0.954 (0.003)} & 0.717 (0.014) & 0.3 (0.01) \\
\multicolumn{1}{c|}{} & \multicolumn{1}{c|}{} & \multicolumn{1}{c|}{} & PLS & 1.00 (0.00) & 0.0 (0.00) & 1.024 (0.028) & 0.243 (0.006) & 0.904 (0.005) & 0.733 (0.015) & 0.2 (0.01) \\
\multicolumn{1}{c|}{} & \multicolumn{1}{c|}{} & \multicolumn{1}{c|}{} & SPC & 4.50 (0.73) & 0.5 (0.17) & 1.163 (0.027) & 0.208 (0.009) & 0.893 (0.007) & 0.731 (0.016) & 1.1 (0.03) \\
\multicolumn{1}{c|}{} & \multicolumn{1}{c|}{} & \multicolumn{1}{c|}{} & SOLAR & 3.00 (0.00) & \textbf{1.0 (0.00)} & \textbf{0.672 (0.006)} & 0.102 (0.005) & \textbf{0.954 (0.003)} & \textbf{0.716 (0.014)} & 1.1 (0.01) \\ \cline{2-11} 
\multicolumn{1}{c|}{} & \multicolumn{1}{c|}{\multirow{12}{*}{5}} & \multicolumn{1}{c|}{\multirow{4}{*}{\begin{tabular}[c]{@{}c@{}}SNR\_X = 0.20\\ SNR\_w = 0.05\end{tabular}}} & PCR & 2.10 (0.28) & 0.0 (0.00) & \textbf{0.510 (0.061)} & 0.261 (0.015) & 0.687 (0.049) & 1.562 (0.039) & 0.1 (0.00) \\
\multicolumn{1}{c|}{} & \multicolumn{1}{c|}{} & \multicolumn{1}{c|}{} & PLS & 1.00 (0.00) & 0.0 (0.00) & 1.122 (0.042) & 0.455 (0.015) & 0.855 (0.017) & 1.572 (0.037) & 0.2 (0.01) \\
\multicolumn{1}{c|}{} & \multicolumn{1}{c|}{} & \multicolumn{1}{c|}{} & SPC & 3.70 (0.86) & 0.0 (0.00) & 1.179 (0.148) & 0.361 (0.033) & 0.782 (0.035) & 1.564 (0.046) & 1.0 (0.02) \\
\multicolumn{1}{c|}{} & \multicolumn{1}{c|}{} & \multicolumn{1}{c|}{} & SOLAR & 5.00 (0.00) & \textbf{1.0 (0.00)} & 0.665 (0.004) & \textbf{0.164 (0.013)} & \textbf{0.891 (0.019)} & \textbf{1.551 (0.040)} & 1.4 (0.01) \\ \cline{3-11} 
\multicolumn{1}{c|}{} & \multicolumn{1}{c|}{} & \multicolumn{1}{c|}{\multirow{4}{*}{\begin{tabular}[c]{@{}c@{}}SNR\_X = 0.01\\ SNR\_w = 0.50\end{tabular}}} & PCR & 3.50 (0.43) & 0.0 (0.00) & 1.945 (0.117) & 0.297 (0.003) & 0.544 (0.013) & 0.598 (0.014) & 0.1 (0.00) \\
\multicolumn{1}{c|}{} & \multicolumn{1}{c|}{} & \multicolumn{1}{c|}{} & PLS & 1.00 (0.00) & 0.0 (0.00) & 1.198 (0.009) & 0.369 (0.005) & \textbf{0.691 (0.010)} & \textbf{0.589 (0.013)} & 0.2 (0.01) \\
\multicolumn{1}{c|}{} & \multicolumn{1}{c|}{} & \multicolumn{1}{c|}{} & SPC & 2.50 (0.97) & 0.0 (0.00) & 1.546 (0.191) & 0.332 (0.005) & 0.625 (0.017) & 0.593 (0.017) & 1.1 (0.02) \\
\multicolumn{1}{c|}{} & \multicolumn{1}{c|}{} & \multicolumn{1}{c|}{} & SOLAR & 1.30 (0.15) & 0.0 (0.00) & \textbf{1.163 (0.063)} & \textbf{0.316 (0.008)} & 0.435 (0.052) & 0.600 (0.016) & 0.6 (0.04) \\ \cline{3-11} 
\multicolumn{1}{c|}{} & \multicolumn{1}{c|}{} & \multicolumn{1}{c|}{\multirow{4}{*}{\begin{tabular}[c]{@{}c@{}}SNR\_X = 0.05\\ SNR\_w = 0.20\end{tabular}}} & PCR & 3.70 (0.30) & 0.3 (0.15) & \textbf{0.953 (0.037)} & 0.169 (0.008) & 0.882 (0.012) & 0.795 (0.022) & 0.1 (0.00) \\
\multicolumn{1}{c|}{} & \multicolumn{1}{c|}{} & \multicolumn{1}{c|}{} & PLS & 1.00 (0.00) & 0.0 (0.00) & 1.143 (0.019) & 0.355 (0.008) & 0.837 (0.009) & 0.797 (0.019) & 0.3 (0.01) \\
\multicolumn{1}{c|}{} & \multicolumn{1}{c|}{} & \multicolumn{1}{c|}{} & SPC & 3.90 (0.87) & 0.1 (0.10) & 1.544 (0.105) & 0.278 (0.008) & 0.831 (0.008) & 0.803 (0.020) & 1.1 (0.03) \\
\multicolumn{1}{c|}{} & \multicolumn{1}{c|}{} & \multicolumn{1}{c|}{} & SOLAR & 5.00 (0.00) & \textbf{1.0 (0.00)} & 1.080 (0.004) & \textbf{0.144 (0.003)} & \textbf{0.920 (0.004)} & \textbf{0.788 (0.020)} & 1.5 (0.02) \\ \hline
\multicolumn{1}{c|}{\multirow{24}{*}{$(1000, 10^5)$}} & \multicolumn{1}{c|}{\multirow{12}{*}{3}} & \multicolumn{1}{c|}{\multirow{4}{*}{\begin{tabular}[c]{@{}c@{}}SNR\_X = 0.20\\ SNR\_w = 0.05\end{tabular}}} & PCR & 3.00 (0.00) & \textbf{1.0 (0.00)} & \textbf{0.360 (0.006)} & 0.065 (0.006) & \textbf{0.972 (0.007)} & \textbf{0.984 (0.013)} & 59.5 (2.45) \\
\multicolumn{1}{c|}{} & \multicolumn{1}{c|}{} & \multicolumn{1}{c|}{} & PLS & 3.70 (0.70) & 0.0 (0.00) & 1.010 (0.134) & 0.743 (0.138) & 0.428 (0.114) & 0.990 (0.014) & 50.9 (2.33) \\
\multicolumn{1}{c|}{} & \multicolumn{1}{c|}{} & \multicolumn{1}{c|}{} & SPC & 5.60 (0.93) & 0.5 (0.17) & 0.388 (0.009) & 0.226 (0.067) & 0.804 (0.079) & 0.989 (0.013) & 554.1 (13.16) \\
\multicolumn{1}{c|}{} & \multicolumn{1}{c|}{} & \multicolumn{1}{c|}{} & SOLAR & 3.00 (0.00) & \textbf{1.0 (0.00)} & \textbf{0.360 (0.006)} & \textbf{0.064 (0.005)} & \textbf{0.972 (0.007)} & \textbf{0.984 (0.012)} & 452.1 (19.62) \\ \cline{3-11} 
\multicolumn{1}{c|}{} & \multicolumn{1}{c|}{} & \multicolumn{1}{c|}{\multirow{4}{*}{\begin{tabular}[c]{@{}c@{}}SNR\_X = 0.01\\ SNR\_w = 0.50\end{tabular}}} & PCR & 3.00 (0.00) & \textbf{1.0 (0.00)} & \textbf{0.387 (0.006)} & \textbf{0.027 (0.002)} & \textbf{0.995 (0.001)} & 0.319 (0.004) & 60.4 (2.68) \\
\multicolumn{1}{c|}{} & \multicolumn{1}{c|}{} & \multicolumn{1}{c|}{} & PLS & 1.00 (0.00) & 0.0 (0.00) & 1.005 (0.016) & 0.150 (0.002) & 0.910 (0.004) & 0.318 (0.004) & 48.1 (1.91) \\
\multicolumn{1}{c|}{} & \multicolumn{1}{c|}{} & \multicolumn{1}{c|}{} & SPC & 3.00 (0.00) & 1.0 (0.00) & 0.593 (0.016) & 0.112 (0.003) & 0.953 (0.003) & \textbf{0.314 (0.004)} & 726.0 (68.78) \\
\multicolumn{1}{c|}{} & \multicolumn{1}{c|}{} & \multicolumn{1}{c|}{} & SOLAR & 3.00 (0.00) & \textbf{1.0 (0.00)} & \textbf{0.387 (0.006)} & \textbf{0.027 (0.001)} & \textbf{0.995 (0.001)} & 0.318 (0.004) & 537.4 (16.57) \\ \cline{3-11} 
\multicolumn{1}{c|}{} & \multicolumn{1}{c|}{} & \multicolumn{1}{c|}{\multirow{4}{*}{\begin{tabular}[c]{@{}c@{}}SNR\_X = 0.05\\ SNR\_w = 0.20\end{tabular}}} & PCR & 3.00 (0.00) & \textbf{1.0 (0.00)} & \textbf{0.363 (0.006)} & \textbf{0.033 (0.003)} & \textbf{0.993 (0.002)} & \textbf{0.493 (0.006)} & 52.1 (3.47) \\
\multicolumn{1}{c|}{} & \multicolumn{1}{c|}{} & \multicolumn{1}{c|}{} & PLS & 4.60 (0.40) & 0.0 (0.00) & 1.327 (0.041) & 0.503 (0.004) & 0.401 (0.011) & 0.495 (0.006) & 50.1 (2.29) \\
\multicolumn{1}{c|}{} & \multicolumn{1}{c|}{} & \multicolumn{1}{c|}{} & SPC & 5.60 (0.93) & 0.5 (0.17) & 0.414 (0.013) & 0.158 (0.037) & 0.856 (0.058) & 0.495 (0.006) & 480.4 (26.60) \\
\multicolumn{1}{c|}{} & \multicolumn{1}{c|}{} & \multicolumn{1}{c|}{} & SOLAR & 3.00 (0.00) & \textbf{1.0 (0.00)} & \textbf{0.363 (0.006)} & 0.034 (0.003) & \textbf{0.993 (0.002)} & \textbf{0.493 (0.006)} & 464.1 (22.26) \\ \cline{2-11} 
\multicolumn{1}{c|}{} & \multicolumn{1}{c|}{\multirow{12}{*}{5}} & \multicolumn{1}{c|}{\multirow{4}{*}{\begin{tabular}[c]{@{}c@{}}SNR\_X = 0.20\\ SNR\_w = 0.05\end{tabular}}} & PCR & 3.60 (0.31) & 0.3 (0.15) & \textbf{0.427 (0.018)} & 0.122 (0.008) & 0.887 (0.014) & 1.144 (0.017) & 40.9 (3.21) \\
\multicolumn{1}{c|}{} & \multicolumn{1}{c|}{} & \multicolumn{1}{c|}{} & PLS & 4.70 (0.40) & 0.3 (0.15) & 1.593 (0.063) & 1.112 (0.009) & 0.222 (0.010) & 1.148 (0.018) & 44.8 (2.44) \\
\multicolumn{1}{c|}{} & \multicolumn{1}{c|}{} & \multicolumn{1}{c|}{} & SPC & 5.20 (0.65) & 0.4 (0.16) & 0.481 (0.012) & 0.355 (0.100) & 0.745 (0.090) & 1.147 (0.019) & 382.6 (9.13) \\
\multicolumn{1}{c|}{} & \multicolumn{1}{c|}{} & \multicolumn{1}{c|}{} & SOLAR & 5.00 (0.00) & \textbf{1.0 (0.00)} & 0.469 (0.005) & \textbf{0.091 (0.009)} & \textbf{0.937 (0.013)} & \textbf{1.142 (0.017)} & 602.2 (3.53) \\ \cline{3-11} 
\multicolumn{1}{c|}{} & \multicolumn{1}{c|}{} & \multicolumn{1}{c|}{\multirow{4}{*}{\begin{tabular}[c]{@{}c@{}}SNR\_X = 0.01\\ SNR\_w = 0.50\end{tabular}}} & PCR & 5.00 (0.00) & \textbf{1.0 (0.00)} & \textbf{0.534 (0.005)} & \textbf{0.039 (0.002)} & \textbf{0.990 (0.001)} & 0.373 (0.004) & 37.8 (2.41) \\
\multicolumn{1}{c|}{} & \multicolumn{1}{c|}{} & \multicolumn{1}{c|}{} & PLS & 1.00 (0.00) & 0.0 (0.00) & 1.126 (0.011) & 0.205 (0.002) & 0.875 (0.004) & \textbf{0.366 (0.005)} & 45.6 (0.96) \\
\multicolumn{1}{c|}{} & \multicolumn{1}{c|}{} & \multicolumn{1}{c|}{} & SPC & 5.00 (0.00) & \textbf{1.0 (0.00)} & 0.899 (0.025) & 0.165 (0.004) & 0.920 (0.003) & \textbf{0.366 (0.005)} & 503.3 (49.52) \\
\multicolumn{1}{c|}{} & \multicolumn{1}{c|}{} & \multicolumn{1}{c|}{} & SOLAR & 5.00 (0.00) & \textbf{1.0 (0.00)} & \textbf{0.534 (0.005)} & \textbf{0.039 (0.002)} & \textbf{0.990 (0.001)} & 0.373 (0.004) & 590.9 (11.54) \\ \cline{3-11} 
\multicolumn{1}{c|}{} & \multicolumn{1}{c|}{} & \multicolumn{1}{c|}{\multirow{4}{*}{\begin{tabular}[c]{@{}c@{}}SNR\_X = 0.05\\ SNR\_w = 0.20\end{tabular}}} & PCR & 4.80 (0.20) & 0.9 (0.10) & \textbf{0.469 (0.010)} & 0.053 (0.007) & 0.977 (0.008) & \textbf{0.572 (0.008)} & 43.2 (4.02) \\
\multicolumn{1}{c|}{} & \multicolumn{1}{c|}{} & \multicolumn{1}{c|}{} & PLS & 4.70 (0.40) & 0.3 (0.15) & 1.751 (0.032) & 0.557 (0.005) & 0.412 (0.009) & 0.575 (0.008) & 43.1 (2.20) \\
\multicolumn{1}{c|}{} & \multicolumn{1}{c|}{} & \multicolumn{1}{c|}{} & SPC & 5.90 (0.59) & 0.4 (0.16) & 0.651 (0.060) & 0.177 (0.028) & 0.894 (0.039) & 0.575 (0.010) & 442.3 (27.54) \\
\multicolumn{1}{c|}{} & \multicolumn{1}{c|}{} & \multicolumn{1}{c|}{} & SOLAR & 5.00 (0.00) & \textbf{1.0 (0.00)} & 0.476 (0.005) & \textbf{0.049 (0.004)} & \textbf{0.983 (0.003)} & \textbf{0.572 (0.008)} & 541.5 (5.58) \\ \hline
\end{tabular}}
\caption{
Representative moderate- and high-dimensional simulation results comparing PCR, PLS, SPC, and SOLAR across varying signal-to-noise regimes and latent ranks. Reported values are means with standard errors across 10 independent replications. Rank recovery proportion denotes the proportion of replications satisfying $\widehat q=q_{\mathrm{true}}$. The $H$-projector error measures recovery of the latent sample-level subspace. Supervised signal RMSE and correlation evaluate recovery of the latent supervised signal $H_{\mathrm{true}}\beta_{\mathrm{true}}$, while test RMSE evaluates out-of-sample prediction accuracy for the response $w$.
}
\label{tab:sim_main}
\end{table}

Table~\ref{tab:sim_main} summarizes representative moderate- and high-dimensional simulation results. Across a broad range of signal regimes, SOLAR achieved competitive or superior supervised signal recovery and predictive performance while simultaneously providing stable adaptive rank estimation. Except for the lower-dimensional setting with weak matrix signal and strong supervised signal, SOLAR achieved nearly perfect rank recovery across all considered scenarios, with recovery accuracy generally improving as both sample size and dimensionality increased. PCR performed strongly in settings where dominant variance directions aligned closely with the supervised latent structure, particularly in higher-dimensional regimes where averaging effects reduced noise contamination in the leading components. Nevertheless, unlike PCR, SOLAR directly incorporates supervision into latent factor estimation while simultaneously performing automatic rank selection through the proposed penalized framework. SPC often exhibited increased computational cost and less stable rank recovery, whereas PLS generally underestimated the latent dimension in more challenging regimes.

\begin{table}[!t]
\centering
\renewcommand{\arraystretch}{1.2}
\resizebox{0.86\columnwidth}{!}{%
\begin{tabular}{ccccccccc}
\hline
$(n,p)$ & $q$ & (SNR\_X, SNR\_w) & $\hat{q}$ & \begin{tabular}[c]{@{}c@{}}Rank recovery\\ prop.\end{tabular} & \begin{tabular}[c]{@{}c@{}}$H$-proj. \\ error\end{tabular} & \begin{tabular}[c]{@{}c@{}}Sup. signal\\ RMSE\end{tabular} & \begin{tabular}[c]{@{}c@{}}Sup. signal \\ corr.\end{tabular} & Time (s) \\ \hline
\multicolumn{1}{c|}{\multirow{6}{*}{$(1000, 10^6)$}} & \multicolumn{1}{c|}{\multirow{3}{*}{3}} & \multicolumn{1}{c|}{(0.20, 0.05)} & 3.00 (0.00) & 1.0 (0.00) & 0.070 (0.007) & 0.043 (0.007) & 0.983 (0.007) & 8134.8 (50.47) \\
\multicolumn{1}{c|}{} & \multicolumn{1}{c|}{} & \multicolumn{1}{c|}{(0.01, 0.50)} & 3.00 (0.00) & 1.0 (0.00) & 0.086 (0.005) & 0.016 (0.002) & 0.998 (0.001) & 7954.2 (84.68) \\
\multicolumn{1}{c|}{} & \multicolumn{1}{c|}{} & \multicolumn{1}{c|}{(0.05, 0.20)} & 3.00 (0.00) & 1.0 (0.00) & 0.072 (0.007) & 0.023 (0.003) & 0.996 (0.002) & 8374.5 (19.42) \\ \cline{2-9} 
\multicolumn{1}{c|}{} & \multicolumn{1}{c|}{\multirow{3}{*}{5}} & \multicolumn{1}{c|}{(0.20, 0.05)} & 5.00 (0.00) & 1.0 (0.00) & 0.101 (0.005) & 0.065 (0.004) & 0.972 (0.005) & 7128.8 (43.10) \\
\multicolumn{1}{c|}{} & \multicolumn{1}{c|}{} & \multicolumn{1}{c|}{(0.01, 0.50)} & 5.00 (0.00) & 1.0 (0.00) & 0.133 (0.004) & 0.024 (0.001) & 0.997 (0.000) & 6967.6 (61.57) \\
\multicolumn{1}{c|}{} & \multicolumn{1}{c|}{} & \multicolumn{1}{c|}{(0.05, 0.20)} & 5.00 (0.00) & 1.0 (0.00) & 0.105 (0.005) & 0.034 (0.002) & 0.993 (0.001) & 7124.4 (58.73) \\ \hline
\multicolumn{1}{c|}{\multirow{6}{*}{$(1000, 10^7)$}} & \multicolumn{1}{c|}{\multirow{3}{*}{3}} & \multicolumn{1}{c|}{(0.20, 0.05)} & 3.00 (0.00) & 1.0 (0.00) & 0.061 (0.006) & 0.042 (0.006) & 0.989 (0.004) & 2973.2 (8.07)$^{\dagger}$ \\
\multicolumn{1}{c|}{} & \multicolumn{1}{c|}{} & \multicolumn{1}{c|}{(0.01, 0.50)} & 3.00 (0.00) & 1.0 (0.00) & 0.067 (0.008) & 0.015 (0.002) & 0.999 (0.001) & 9641.7 (497.34)$^{\dagger}$ \\
\multicolumn{1}{c|}{} & \multicolumn{1}{c|}{} & \multicolumn{1}{c|}{(0.05, 0.20)} & 3.00 (0.00) & 1.0 (0.00) & 0.065 (0.008) & 0.022 (0.004) & 0.996 (0.001) & 9102.4 (524.20)$^{\dagger}$ \\ \cline{2-9} 
\multicolumn{1}{c|}{} & \multicolumn{1}{c|}{\multirow{3}{*}{5}} & \multicolumn{1}{c|}{(0.20, 0.05)} & 5.00 (0.00) & 1.0 (0.00) & 0.088 (0.007) & 0.079 (0.007) & 0.962 (0.004) & 2943.7 (12.66)$^{\dagger}$ \\
\multicolumn{1}{c|}{} & \multicolumn{1}{c|}{} & \multicolumn{1}{c|}{(0.01, 0.50)} & 5.00 (0.00) & 1.0 (0.00) & 0.093 (0.006) & 0.027 (0.002) & 0.996 (0.001) & 6761.8 (189.63)$^{\dagger}$ \\
\multicolumn{1}{c|}{} & \multicolumn{1}{c|}{} & \multicolumn{1}{c|}{(0.05, 0.20)} & 5.00 (0.00) & 1.0 (0.00) & 0.089 (0.007) & 0.041 (0.004) & 0.989 (0.001) & 2423.7 (8.28)$^{\dagger}$ \\ \hline
\end{tabular}}
\caption{
Ultra-high-dimensional simulation results for SOLAR at \(p=10^6\) and \(p=10^7\). Reported values are means with standard errors across 10 independent replications. To emphasize scalability and latent recovery under extreme dimensionality, these experiments were performed without train-test splitting. Runtime values marked with \(^{\dagger}\) were obtained using 40-thread multi-threaded execution, whereas all other computations used a single computational core.
}
\label{tab:sim_ultrahd}
\end{table}

To assess scalability, we additionally considered ultra-high-dimensional settings with $(n,p)\in\{(1000,10^6),(1000,10^7)\}$. Due to the extreme memory and computational requirements associated with fitting and cross-validating competing methods at these dimensions, only SOLAR was evaluated in these experiments. Since the primary focus was scalability and latent recovery rather than prediction, train-test splitting was not performed in these settings. Instead, evaluation focused on rank recovery, latent subspace recovery, supervised signal recovery, and runtime behavior. Table~\ref{tab:sim_ultrahd} summarizes the ultra-high-dimensional experiments. SOLAR consistently recovered the correct latent rank across all considered settings, while maintaining stable latent-space recovery and strong supervised signal agreement even at $p=10^7$. These results demonstrate that the proposed framework remains computationally feasible and statistically stable in dimensions substantially beyond the range typically considered in supervised latent factor modeling.
\section{Developmental epigenetic aging heterogeneity and\\ 
supervised latent methylation structure}
\label{sec:realdata_results}
We next investigate the scientific questions raised in Section~\ref{sec:realdata} using longitudinal methylation profiles from the GUSTO early-childhood cohort \citep{leroy2025longitudinal}. Genome-wide DNAm was measured using the Illumina Infinium Methylation EPIC platform, yielding $p=865{,}859$ CpG probes across $n=1051$ methylation samples collected approximately at 3, 9, 48, and 72 months of age. The resulting data therefore represent an ultra-high-dimensional developmental methylation setting ($p \gg n$) spanning infancy through early childhood, with substantial dependence and correlation across CpG sites. Because the primary objective of the present analysis is recovery of supervised developmental methylation structure associated with biological-aging heterogeneity, rather than subject-specific longitudinal trajectory estimation, each methylation profile is treated as a high-dimensional developmental snapshot while developmental patterns are subsequently examined through the inferred latent factors across age groups.

As discussed in Section~\ref{sec:realdata}, the primary scientific interest concerns biological aging variation beyond chronological developmental progression itself. Consequently, subsequent analyses focus on residualized DNAm age measures rather than raw DNAm age estimates. Operationally, these outcomes quantify deviation from expected age-related methylation patterns after accounting for chronological age and are widely used in the biological-aging literature to characterize heterogeneous aging-related processes beyond normative maturation \citep{horvath2013dna,levine2018epigenetic}. Although the Horvath, Levine, and Hannum clocks are themselves constructed using very small subsets of CpG measurements, the present analysis does not aim to reconstruct chronological age prediction models. Rather, the goal is to identify broader supervised latent methylation structure associated with residual developmental biological-aging heterogeneity beyond age-expected methylation trends. Consequently, the proposed framework uses the full methylation profile to recover coordinated developmental methylation programs associated with residualized developmental biological age rather than merely reproducing the original clock constructions, while simultaneously allowing biologically relevant CpGs not originally included in the corresponding epigenetic clocks to contribute to the inferred supervised latent structure.

\subsection{Exploratory developmental epigenetic aging patterns}
\begin{figure}[!t]
\centering
\includegraphics[width=0.99\textwidth]{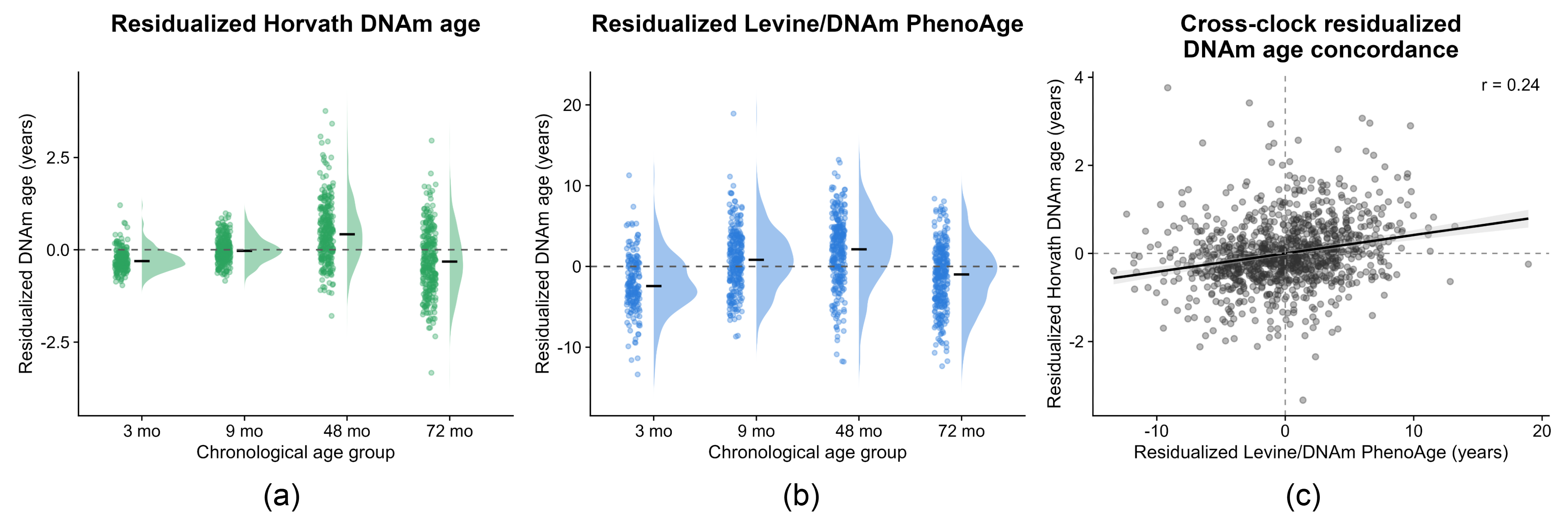}
\caption{
Exploratory characterization of residualized DNAm age across early childhood.
(a) Residualized Horvath DNAm age across developmental age groups (3, 9, 48, and 72 months). 
(b) Residualized Levine/DNAm PhenoAge across developmental age groups. 
Points represent individual subjects, violin densities summarize within-group distributions, and horizontal black segments denote group means. Dashed horizontal lines indicate zero residualized DNAm age. 
(c) Cross-clock concordance between residualized Horvath and Levine DNAm age measures. The fitted regression line and corresponding correlation illustrate moderate but nontrivial agreement between the two developmental biological-aging measures.
}
\label{fig:residual_heterogeneity}
\end{figure}

We first addressed Research Question 1 by investigating whether early-life methylation profiles exhibit heterogeneous developmental biological aging patterns beyond chronological age alone. Figure~\ref{fig:residual_heterogeneity} summarizes the resulting residualized DNAm age patterns for the Horvath and Levine clocks across developmental age groups. Residualized Horvath DNAm age remained comparatively well-centered across childhood, although substantial within-group heterogeneity persisted throughout development (Figure~\ref{fig:residual_heterogeneity}a). In contrast, residualized Levine/DNAm PhenoAge exhibited substantially broader dispersion and markedly larger variability across age groups (Figure~\ref{fig:residual_heterogeneity}b), suggesting that the Levine-derived residualized measure may reflect broader physiological and phenotypic heterogeneity beyond the comparatively developmentally calibrated Horvath measure. These residual heterogeneity patterns also parallel the broader developmental calibration differences observed earlier in Figures~\ref{fig:motivation}(d)--(g), where Horvath DNAm age demonstrated the strongest developmental coherence with chronological age, Levine exhibited intermediate variability, and Hannum displayed comparatively unstable behavior in this pediatric setting. Consequently, residualized Horvath DNAm age was selected as the primary supervised outcome for the main SOLAR analysis, Levine analyses were retained as secondary sensitivity analyses and reported in the Supplementary Section C, whereas Hannum was not pursued further.

Despite their differing developmental behaviors, residualized Horvath and Levine DNAm age remained positively associated (Figure~\ref{fig:residual_heterogeneity}c), with moderate cross-clock concordance ($r=0.24$). Collectively, these findings suggest that developmental methylation heterogeneity cannot be fully characterized through chronological age alone and motivate the need for supervised low-dimensional methylation structure capable of capturing heterogeneous biological-aging variation during childhood.

\subsection{Supervised latent methylation structure}
\begin{figure}[!t]
\centering
\includegraphics[width=0.99\textwidth]{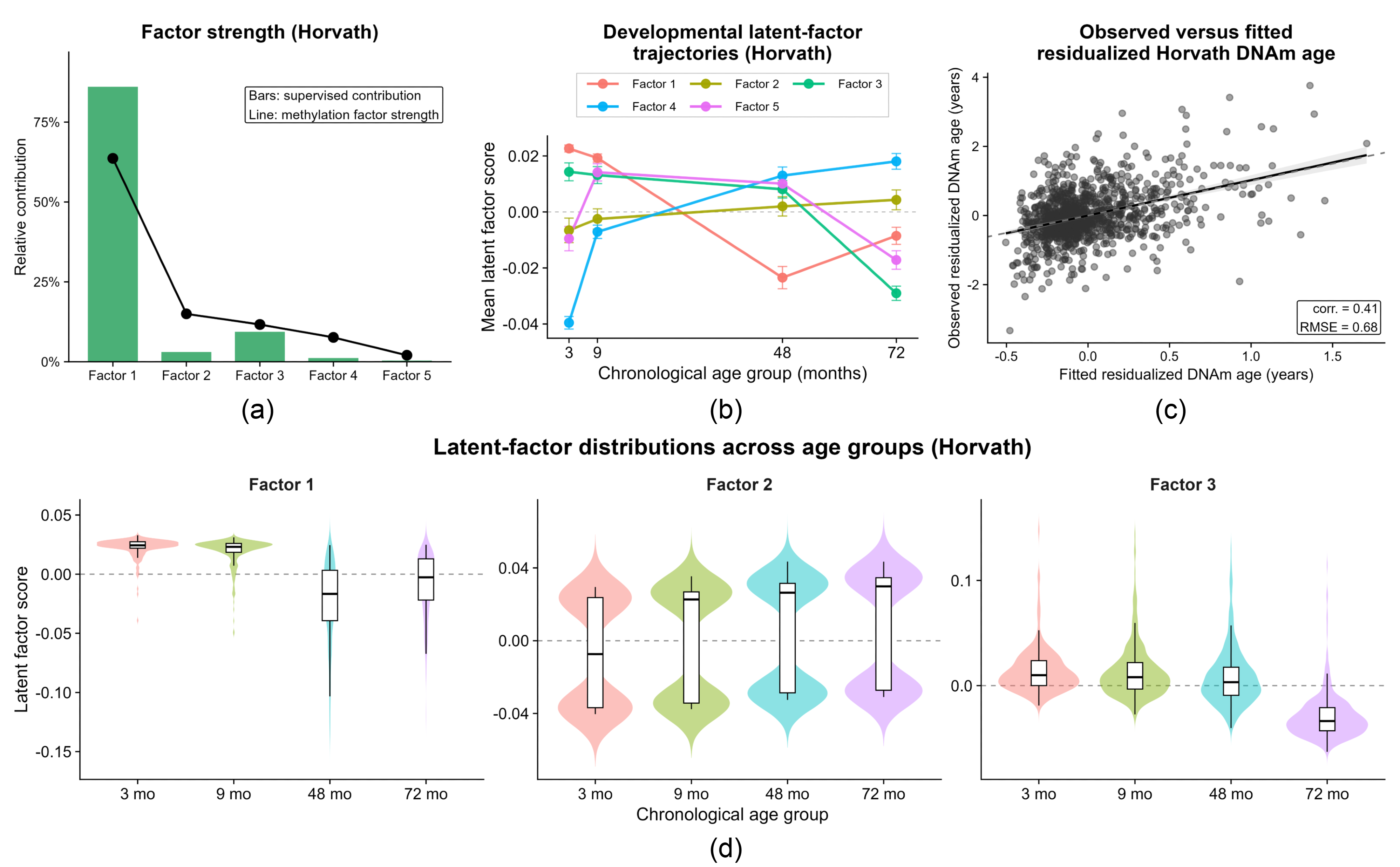}
\caption{
SOLAR latent-factor characterization for residualized Horvath DNAm age.
(a) Relative supervised contributions of the inferred latent methylation factors together with their corresponding latent-factor strengths derived from the singular-value structure. 
(b) Mean developmental trajectories of the inferred latent factors across chronological age groups, with error bars denoting standard errors. 
(c) Observed versus fitted residualized Horvath DNAm age values from the fitted SOLAR model. 
(d) Distribution of selected latent-factor scores across developmental age groups, illustrating substantial within-group heterogeneity and evolving developmental methylation structure.
}
\label{fig:horvath_latent}
\end{figure}

To address Research Question 2, we next applied SOLAR using residualized Horvath DNAm age as the primary supervised outcome. Figure~\ref{fig:horvath_latent}(a) summarizes the inferred latent-factor structure through two complementary quantities: the latent methylation factor strengths based on the squared singular values $d_k^2$, and the corresponding supervised contributions $d_k^2|\beta_k|$, which additionally incorporate association with developmental biological-aging variation. The first supervised methylation factor clearly dominated the latent representation, accounting for the majority of supervised developmental aging heterogeneity, whereas the remaining factors contributed more moderate but non-negligible structure. Importantly, these latent factors were not estimated merely to maximize unsupervised methylation variance; rather, they represent coordinated methylation structure specifically associated with developmental biological-aging heterogeneity.

The inferred latent factors additionally exhibited distinct developmental patterns across childhood (Figure~\ref{fig:horvath_latent}(b)). Several factors demonstrated nonlinear temporal trajectories, including directional reversals between infancy and later childhood, suggesting that developmental biological aging heterogeneity may arise through multiple evolving methylation processes rather than a single monotone developmental mechanism. Figure~\ref{fig:horvath_latent}(d) further illustrates substantial within-age-group heterogeneity in the latent-factor distributions. In particular, Factors 1 and 3 exhibited progressively broader dispersion at later developmental stages, whereas Factor 2 displayed persistent bimodal separation across age groups, potentially reflecting heterogeneous developmental substructure not fully explained by chronological age alone. Collectively, these findings support the presence of heterogeneous developmental methylation structure beyond normative age-related maturation.

The observed-versus-fitted relationship in Figure~\ref{fig:horvath_latent}(c) demonstrated moderate but nontrivial agreement between the inferred supervised latent structure and observed residualized Horvath DNAm age ($r=0.41$). Because the outcome had already been residualized with respect to chronological age, the inferred latent factors reflect developmental methylation variation beyond normative age-related maturation rather than simple rediscovery of chronological age itself. Figure~\ref{fig:horvath_cpg}(c) additionally demonstrated stable low-dimensional structure under bootstrap resampling, with the selected latent rank concentrated primarily around five to six factors despite the ultra-high-dimensional methylation space.

\subsection{CpG importance structure and biological interpretation}
To address Research Question 3, we next investigated which methylation signatures contributed most strongly to developmental biological-aging heterogeneity and whether these signatures remained biologically coherent across epigenetic-aging measures. Building upon the fitted SOLAR decomposition $X \approx H D V^\top$, where $D=\mathrm{diag}(d_1,\ldots,d_q)$ contains latent factor strengths and $V=(v_{jk})$ contains the CpG loading structure, we quantified CpG-specific contribution using supervised importance scores that combine latent-factor dominance with association to developmental biological-aging variation through the supervised coefficients $\beta=(\beta_1,\ldots,\beta_q)^\top$. Specifically, we define
\[
\mathrm{Imp}_j
=
\sum_{k=1}^q
d_k^2 |\beta_k| v_{jk}^2,
\qquad
\mathrm{RIS}_j
=
\frac{\mathrm{Imp}_j}
{\max_{\ell} \mathrm{Imp}_{\ell}},
\]

\noindent where $\mathrm{RIS}_j$ denotes the supervised relative importance score (RIS), with larger $\mathrm{RIS}_j \in [0,1]$ values indicating stronger relative contribution to supervised developmental biological-aging heterogeneity.

\begin{figure}[!t]
\centering
\includegraphics[width=0.99\textwidth]{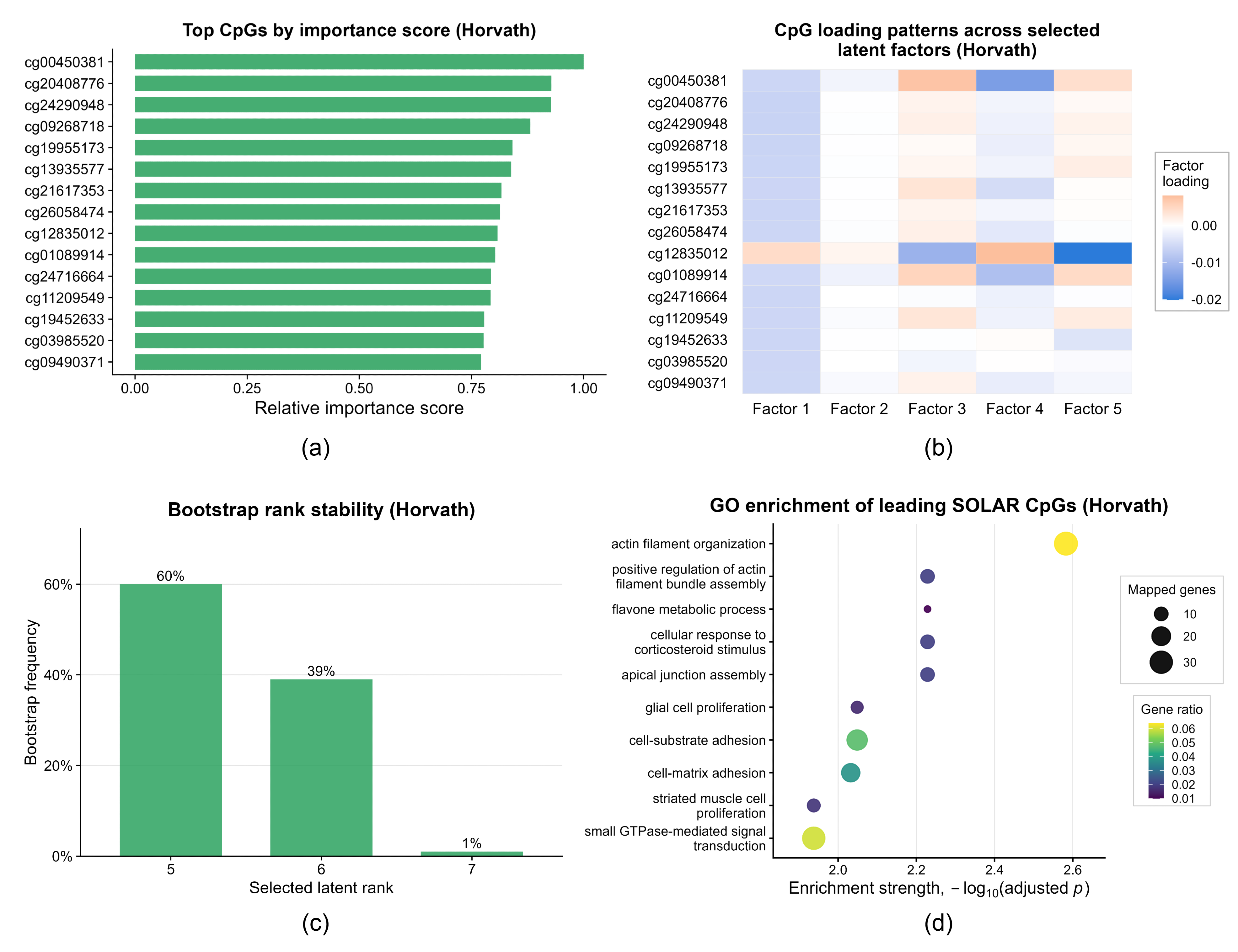}
\caption{
CpG-level interpretation and biological characterization of the SOLAR latent methylation structure for residualized Horvath DNAm age. 
(a) SOLAR identified top 15 CpGs and their relative importance scores. 
(b) Loading patterns of identified top 15 CpGs across the inferred latent factors. 
(c) Bootstrap stability of the selected latent rank across resampled datasets. 
(d) Gene Ontology (GO) enrichment analysis for genes mapped from top 1000 leading SOLAR CpGs, demonstrating enrichment for developmental and cellular regulatory pathways associated with biological aging heterogeneity.
}
\label{fig:horvath_cpg}
\end{figure}

Figure~\ref{fig:horvath_cpg}(a) displays the top 15 ranked CpGs according to the resulting relative importance scores. Several leading CpGs demonstrated highly concentrated loading patterns across specific latent factors (Figure~\ref{fig:horvath_cpg}(b)); for example, cg12835012 (rank 9) exhibited strong positive loading on Factor 4 together with comparatively weaker contributions across the remaining latent dimensions. These patterns suggest that the inferred developmental methylation programs are partially factor-specific rather than diffusely distributed across all latent dimensions. In particular, multiple highly ranked CpGs loaded strongly on the dominant latent factor identified earlier in Figure~\ref{fig:horvath_latent}(a), further supporting the existence of structured supervised developmental methylation programs associated with biological-aging heterogeneity.

To investigate broader biological coherence of the inferred methylation signatures, we additionally performed Gene Ontology enrichment analysis using genes mapped from the top 1,000 CpGs ranked according to the proposed supervised importance score. Enrichment analysis evaluates whether biologically related functional pathways are overrepresented among the highly ranked CpGs relative to background expectation. Biological Process ontology terms were assessed using Benjamini--Hochberg false discovery rate correction, and Figure~\ref{fig:horvath_cpg}(d) demonstrated significant FDR-adjusted enrichment for pathways related to actin filament organization, cell adhesion, glial proliferation, apical junction assembly, and small GTPase-mediated signaling. Several of these pathways have previously been implicated in developmental regulation, cellular differentiation, inflammatory signaling, and aging-related cellular maintenance processes \citep{jylhava2017biological}. Although the present analysis is not intended to establish causal biological mechanisms, the observed enrichment patterns suggest that the supervised latent methylation structure recovered by SOLAR reflects biologically coherent developmental programs rather than purely statistical artifacts. Additional details of the GO enrichment results, including GO identifiers, enrichment ratios, gene and background ratios, adjusted (p)-values, mapped gene counts, mapped CpG counts, and contributing genes, are provided in Supplementary Table~S5.

Parallel analyses using residualized Levine/DNAm PhenoAge yielded qualitatively similar but biologically broader latent structure (Supplementary Figure~S1), with greater residual variability, more diffuse developmental latent trajectories, and stronger concentration of several leading CpGs within a dominant latent factor relative to the Horvath analyses. The top 50 CpGs identified under both residualized Horvath and Levine/DNAm PhenoAge, together with their ranks, supervised relative importance scores (RIS), dominant latent factors, and annotated genes, are reported in Supplementary Tables~S3--S4. Notably, two CpGs appeared in the top-50 lists for both clocks: cg00450381 ranked first for both residualized Horvath and Levine/DNAm PhenoAge (RIS = 1.000 in both analyses; annotated to \textit{EYS}), while cg17382258 ranked 36th and 37th, with RIS values 0.703 and 0.701, respectively. Supplementary Table~S2 additionally summarizes an out-of-sample comparison based on a common 80\%/20\% training--test split under a fixed latent dimension ($q=5$), thereby allowing predictive behavior across methods to be compared under a shared low-rank specification. Although PLS achieved slightly stronger predictive performance in this setting, SOLAR demonstrated comparable predictive behavior to PCR while additionally providing stable supervised latent decomposition, interpretable CpG loading structure, and biologically coherent downstream enrichment characterization. Collectively, these findings suggest that residualized DNAm age variation is associated with both shared and clock-specific supervised methylation structure, although the clock-specific components may reflect a combination of calibration error, measurement bias, and potentially distinct biological signal.

\section{Discussion}\label{sec:discussion}

This work develops a supervised orthogonal low-rank framework for extracting interpretable latent methylation structure associated with developmental epigenetic aging in ultra-high-dimensional settings. Across  two simulation studies and real-world testing in the GUSTO developmental methylation cohort, SOLAR recovered biologically coherent low-dimensional methylation structure while remaining computationally feasible at methylation-array scale. In the case-study analyses, the proposed framework identified substantial developmental heterogeneity beyond chronological age alone and revealed partially shared yet clock-specific methylation signatures across residualized Horvath and Levine measures.

An important practical feature of the proposed framework is its computational and memory scalability in ultra-high-dimensional settings. Modern methylation arrays routinely contain hundreds of thousands to millions of CpGs, making memory usage a major computational bottleneck in practice. Table~\ref{tab:memory_table} summarizes runtime and memory behavior of SOLAR for single end-to-end analyses of synthetic methylation matrices with \(p=10^6\) CpGs across multiple sample sizes. Importantly, these scalability experiments were conducted on a standard desktop workstation (32 GB RAM, 12 CPU cores, Intel i7-12700 at 2.10 GHz) without requiring distributed or cluster computing infrastructure. The methylation matrices were stored in single precision, reducing storage and memory overhead while preserving stable numerical behavior throughout optimization. Even for \(n=2000\) and \(p=10^6\), where the methylation matrix itself occupied approximately 6.9 GB on disk, the complete end-to-end workflow, including data loading, optimization, and result generation, remained executable within approximately 25 GB RAM under the above desktop configuration. This scalability is facilitated by several deliberately memory-aware implementation strategies, including (i) Gram-matrix spectral initialization together with partial low-rank eigen-decomposition rather than direct decomposition of the full methylation matrix, (ii) trace-based objective evaluation avoiding explicit reconstruction of \(HDV^\top\), and (iii) single-precision binary matrix storage for ultra-high-dimensional methylation arrays. Moreover, the demonstrated scalability of SOLAR up to \(p=10^7\) under cluster-computing settings places the proposed framework beyond the dimensional range of conventional array-based methylation analyses and supports its use in substantially larger feature-selected epigenomic applications. Whole-genome bisulfite sequencing and single-cell methylation studies introduce additional scale and design complexities, including broader methylation contexts and single-base resolution \citep{kremer2024analyzing}; these settings provide natural opportunities for future extensions of SOLAR. Collectively, these considerations show that SOLAR is computationally tractable for the present EPIC-array analysis and scalable to ultra-high-dimensional methylation studies beyond standard array-scale data.

\begin{table}[!t]
\centering
\renewcommand{\arraystretch}{1.2}
\resizebox{0.85\columnwidth}{!}{%
\begin{tabular}{cccccccc}
\hline
$p$ & $n$ & Hardware & \begin{tabular}[c]{@{}c@{}}Size\\ of $X$\end{tabular} & \begin{tabular}[c]{@{}c@{}}Precision\\ of $X$\end{tabular} & \begin{tabular}[c]{@{}c@{}}Runtime\\ (s)\end{tabular} & \begin{tabular}[c]{@{}c@{}}Peak MATLAB\\ memory \\ footprint\end{tabular} & \begin{tabular}[c]{@{}c@{}}Observed\\ peak RAM\end{tabular} \\ \hline
\multirow{3}{*}{$10^6$} & 500 & \multirow{3}{*}{\begin{tabular}[c]{@{}c@{}}32 GB desktop, 12 CPU cores, \\ Intel i7-12700 (2.10 GHz)\end{tabular}} & 1.73 GB & single & 147 & 5.44 GB & $\sim$ 7 GB \\
 & 1000 &  & 3.46 GB & single & 275 & 7.31 GB & $\sim$ 9 GB \\
 & 2000 &  & 6.92 GB & single & 553 & 11.00 GB & $\sim$ 25 GB \\ \hline
\end{tabular}}
\caption{Runtime and memory scalability of SOLAR for ultra-high-dimensional methylation matrices with \(p=10^6\) CpGs under varying sample sizes. Experiments were conducted on a standard desktop workstation (32 GB RAM, 12 CPU cores, Intel i7-12700 at 2.10 GHz) using single-precision matrix storage. Peak MATLAB memory footprint denotes the maximum MATLAB-reported memory allocation during optimization, whereas observed peak RAM denotes total end-to-end system memory usage during execution.}
\label{tab:memory_table}
\end{table}

The simulation studies additionally revealed that, in certain high-dimensional regimes where dominant variance directions aligned closely with the supervised signal, SOLAR and classical principal component regression (PCR) exhibited similar latent recovery and predictive behavior. This observation suggests that, under sufficiently strong low-rank signal, leading unsupervised singular directions of \(X\) may already contain substantial supervised information regarding the response. Nevertheless, unlike PCR, SOLAR explicitly incorporates supervision during latent-factor estimation while simultaneously performing adaptive rank selection and providing interpretable CpG-level loading structure and downstream biological characterization.

Several methodological directions remain open for future investigation. Although the proposed dimension-adaptive penalty scaling \(\kappa(n,p)\) demonstrated stable empirical behavior across a broad range of dimensional regimes, the present work does not establish optimality properties for the corresponding penalty formulation. Further theoretical investigation of asymptotically optimal penalty scaling under jointly diverging \((n,p)\) regimes therefore remains of interest. Extensions toward sparse loading structures, generalized supervised outcomes, and longitudinal latent-process formulations may further broaden the applicability of the framework.

From a scientific perspective, the analyses support the growing view that developmental biological aging cannot be fully characterized by chronological age alone and may instead involve heterogeneous low-dimensional methylation programs evolving throughout early childhood. Residualized Horvath DNAm age exhibited comparatively stable developmental calibration and more coherent latent trajectories, whereas residualized Levine/DNAm PhenoAge demonstrated broader residual heterogeneity and more diffuse latent structure. Overall, the proposed SOLAR framework provides a scalable and interpretable supervised low-rank approach for studying developmental epigenetic aging and other ultra-high-dimensional supervised latent-structure problems.

\begin{center}
{\large\bf SUPPLEMENTARY MATERIAL}
\end{center}
\begin{description}
\item[Supplementary text:] Supplementary material is provided as a separate pdf document.

\item[Code and data:] Reproducible code for the simulation studies and SOLAR case-study analyses, demos to fit them to any similarly structured dataset, are made available on GitHub at \href{https://github.com/priyamdas2/SOLAR}{https://github.com/priyamdas2/SOLAR}. The dataset is obtained from the Gene Expression Omnibus (GEO) under accession GSE254135. 
\end{description}
\section*{Disclosure statement}
The authors report there are no competing interests to declare.
\section*{Funding}
PD is supported in part by NIH P30CA016059. KAA is supported in part by NIH R01MH124981.

\bibliographystyle{plainnat}
\bibliography{MM-MC}

\end{document}


\def\spacingset#1{\renewcommand{\baselinestretch}%
{#1}\small\normalsize} \spacingset{1}


\if1\blind
{
  \title{\bf Supplementary Material for ``Supervised Low-Rank Structure Discovery for Developmental Epigenetic Aging in Ultra-High-Dimensional DNA Methylation Data''}
  \author{%
\begin{tabular}{c}
Priyam Das\\
{\normalsize Department of Biostatistics, Virginia Commonwealth University}\\[2mm]
Jiyeon Song\\
{\normalsize Department of Biostatistics, University of Michigan}\\[2mm]
Lathika Mohanraj\\
{\normalsize Department of Adult Health and Nursing Systems, Virginia Commonwealth University}\\[2mm]
Karolina A. Aberg\\
{\normalsize Center for Biomarker Research and Precision Medicine,}\\
{\normalsize Virginia Commonwealth University}\\[2mm]
Yi Li\\
{\normalsize Department of Biostatistics, University of Michigan}\\[2mm]
and\\[1mm]
Subharup Guha\\
{\normalsize Department of Biomedical Data Science, Dartmouth College}
\end{tabular}%
}
\date{}
  \maketitle
} \fi

\if0\blind
{
  \bigskip
  \bigskip
  \bigskip
  \begin{center}
    {\large\bf Supplementary Material for ``Supervised Low-Rank Structure Discovery for Developmental Epigenetic Aging in Ultra-High-Dimensional DNA Methylation Data''}
\end{center}
  \medskip
} \fi

\bigskip

\tableofcontents

\newpage
\spacingset{1.45} 

\section{Detailed Simulation Design and Implementation}
\label{sec:supp_simulation}

This section provides full implementation details for the simulation studies reported in the main manuscript. All simulations were implemented in MATLAB R2024a and run on a Rocky Linux 9 high-performance computing cluster with AMD EPYC processors. Moderate- and high-dimensional simulations were run using 256 GB RAM and single-thread execution. Ultra-high-dimensional SOLAR simulations were run on the same cluster using 494 GB RAM and 40-thread multithreaded execution.

\subsection{Data-generating mechanism}

For each simulation setting, data were generated from the supervised low-rank model
\[
X = H_0D_0V_0^\top + E,
\qquad
w = H_0\beta_0 + \varepsilon,
\]
where \(X\in\mathbb{R}^{n\times p}\) is the high-dimensional predictor matrix, \(w\in\mathbb{R}^n\) is the supervised response, \(H_0\in\mathbb{R}^{n\times q_0}\) and \(V_0\in\mathbb{R}^{p\times q_0}\) are the true left and right latent factor matrices, \(D_0=\mathrm{diag}(d_1,\ldots,d_{q_0})\) is the true diagonal loading matrix, and \(\beta_0\in\mathbb{R}^{q_0}\) is the true supervised coefficient vector. The true rank was \(q_0\in\{3,5\}\).

For each replication, \(H_0\) and \(V_0\) were generated independently by applying QR decompositions to independent standard Gaussian matrices of dimensions \(n\times q_0\) and \(p\times q_0\), respectively. The entries of \(E\) were generated independently as \(N(0,\sigma^2)\), with \(\sigma^2=0.05^2\). The supervised coefficient vector was generated once for each simulation scenario as \(\beta_0=5z\), where \(z\sim N(0,I_{q_0})\). The response noise variance \(\tau^2\) was chosen to match the target supervised signal-to-noise ratio.

The signal strength in \(X\) was controlled through
\[
\mathrm{SNR}_X
=
\frac{\|H_0D_0V_0^\top\|_F^2}{np\sigma^2}
=
\frac{\|D_0\|_F^2}{np\sigma^2},
\]
and the supervised signal strength was controlled through
\[
\mathrm{SNR}_w
=
\frac{\|H_0\beta_0\|_2^2}{n\tau^2}.
\]
We considered three representative signal regimes:
\[
(\mathrm{SNR}_X,\mathrm{SNR}_w)
\in
\{(0.20,0.05),\ (0.01,0.50),\ (0.05,0.20)\},
\]
corresponding to high matrix signal with low supervised signal, low matrix signal with high supervised signal, and moderate matrix/supervised signal, respectively.

The diagonal entries of \(D_0\) were generated to satisfy two constraints: they were separated relative to the Gaussian noise spectral edge, and their total squared magnitude matched the desired \(\mathrm{SNR}_X\). Specifically, let
\[
b_{n,p}=\sigma(\sqrt n+\sqrt p)
\]
denote the approximate upper spectral edge of an \(n\times p\) Gaussian noise matrix. For \(q_0\) latent factors, let \(\alpha_k\) denote an equally spaced decreasing sequence between \(\alpha_{\max}=1.5\) and \(\alpha_{\min}=1.15\), and set preliminary singular values
\[
s_k=b_{n,p}\alpha_k,\qquad k=1,\ldots,q_0.
\]
These values were then rescaled by
\[
c=
\left\{
\frac{\mathrm{SNR}_X\,np\sigma^2}{\sum_{k=1}^{q_0}s_k^2}
\right\}^{1/2},
\]
and the final diagonal entries were
\[
d_k=cs_k,\qquad k=1,\ldots,q_0.
\]
This construction ensures that
\[
\sum_{k=1}^{q_0}d_k^2
=
\mathrm{SNR}_X\,np\sigma^2,
\]
while preserving a detectable singular-value profile relative to the noise bulk edge.

For each scenario, \(N_{\mathrm{rep}}=10\) independent replications were generated. For moderate- and high-dimensional simulations, the generated datasets were saved and reused across methods so that PCR, PLS, SPC, and SOLAR were evaluated on identical replications.

\subsection{Simulation scenarios}

The moderate- and high-dimensional comparison study considered
\[
(n,p)\in
\{(500,1000),\ (500,10000),\ (1000,10000),\ (1000,100000)\},
\]
with \(q_0\in\{3,5\}\) and the three signal regimes described earlier. For these settings, all methods were evaluated using the same 80/20 train-test split. Representative results for the intermediate-dimensional settings \((500,10^4)\) and \((1000,10^4)\), omitted from the main paper due to space constraints, are reported in Table~\ref{tab:sim_supp}. Overall, the supplementary results exhibit trends consistent with those reported in the main paper. SOLAR maintained highly stable rank recovery across most settings while achieving competitive or superior supervised signal recovery and prediction accuracy. In higher-signal regimes, PCR occasionally achieved performance comparable to SOLAR, particularly when dominant variance directions aligned closely with the supervised latent structure. However, SOLAR generally provided more stable adaptive rank estimation across varying signal configurations without requiring cross-validation-based rank tuning. SPC frequently exhibited larger computational cost and more variable rank recovery, whereas PLS tended to underestimate the latent dimension in more challenging regimes.

The ultra-high-dimensional scalability study considered
\[
(n,p)\in\{(1000,10^6),(1000,10^7)\}.
\]
Only SOLAR was evaluated in these settings. To avoid unnecessary disk usage, datasets were generated directly inside each replication rather than saved as external data files. Train-test splitting was not performed in the ultra-high-dimensional study; the goal of this experiment was to assess rank recovery, latent recovery, supervised signal recovery, and computational feasibility under extreme dimensionality.

\begin{table}[!t]
\centering
\renewcommand{\arraystretch}{1.2}
\resizebox{0.99\columnwidth}{!}{%
\begin{tabular}{ccccccccccc}
\hline
$(n,p)$ & $q$ & SNR & Method & $\hat{q}$ & \begin{tabular}[c]{@{}c@{}}Rank recovery\\ prop.\end{tabular} & \begin{tabular}[c]{@{}c@{}}$H$-proj. \\ error\end{tabular} & \begin{tabular}[c]{@{}c@{}}Sup. signal\\ RMSE\end{tabular} & \begin{tabular}[c]{@{}c@{}}Sup. signal \\ corr.\end{tabular} & \begin{tabular}[c]{@{}c@{}}Test RMSE\\ ($w$)\end{tabular} & Time (s) \\ \hline
\multicolumn{1}{c|}{\multirow{24}{*}{$(500, 10^4)$}} & \multicolumn{1}{c|}{\multirow{12}{*}{3}} & \multicolumn{1}{c|}{\multirow{4}{*}{\begin{tabular}[c]{@{}c@{}}SNR\_X = 0.20\\ SNR\_w = 0.05\end{tabular}}} & PCR & 2.40 (0.27) & 0.6 (0.16) & \textbf{0.320 (0.017)} & 0.162 (0.029) & 0.812 (0.062) & 1.437 (0.027) & 2.5 (0.45) \\
\multicolumn{1}{c|}{} & \multicolumn{1}{c|}{} & \multicolumn{1}{c|}{} & PLS & 2.70 (0.76) & 0.1 (0.10) & 1.355 (0.104) & 0.730 (0.185) & 0.625 (0.123) & 1.455 (0.028) & 2.4 (0.19) \\
\multicolumn{1}{c|}{} & \multicolumn{1}{c|}{} & \multicolumn{1}{c|}{} & SPC & 3.70 (1.00) & 0.4 (0.16) & 0.561 (0.068) & 0.251 (0.050) & 0.855 (0.045) & 1.437 (0.027) & 9.0 (0.20) \\
\multicolumn{1}{c|}{} & \multicolumn{1}{c|}{} & \multicolumn{1}{c|}{} & SOLAR & 3.00 (0.00) & \textbf{1.0 (0.00)} & 0.356 (0.006) & \textbf{0.115 (0.010)} & \textbf{0.955 (0.009)} & \textbf{1.433 (0.026)} & 21.8 (1.47) \\ \cline{3-11} 
\multicolumn{1}{c|}{} & \multicolumn{1}{c|}{} & \multicolumn{1}{c|}{\multirow{4}{*}{\begin{tabular}[c]{@{}c@{}}SNR\_X = 0.01\\ SNR\_w = 0.50\end{tabular}}} & PCR & 3.00 (0.00) & \textbf{1.0 (0.00)} & 0.619 (0.005) & 0.086 (0.002) & \textbf{0.966 (0.002)} & 0.480 (0.008) & 2.0 (0.21) \\
\multicolumn{1}{c|}{} & \multicolumn{1}{c|}{} & \multicolumn{1}{c|}{} & PLS & 1.00 (0.00) & 0.0 (0.00) & 1.120 (0.006) & 0.289 (0.004) & 0.826 (0.006) & 0.469 (0.009) & 2.4 (0.21) \\
\multicolumn{1}{c|}{} & \multicolumn{1}{c|}{} & \multicolumn{1}{c|}{} & SPC & 3.70 (0.33) & 0.4 (0.16) & 1.110 (0.007) & 0.239 (0.004) & 0.868 (0.006) & \textbf{0.468 (0.009)} & 9.5 (0.17) \\
\multicolumn{1}{c|}{} & \multicolumn{1}{c|}{} & \multicolumn{1}{c|}{} & SOLAR & 3.00 (0.00) & \textbf{1.0 (0.00)} & \textbf{0.618 (0.005)} & \textbf{0.085 (0.002)} & 0.967 (0.002) & 0.479 (0.008) & 22.6 (0.92) \\ \cline{3-11} 
\multicolumn{1}{c|}{} & \multicolumn{1}{c|}{} & \multicolumn{1}{c|}{\multirow{4}{*}{\begin{tabular}[c]{@{}c@{}}SNR\_X = 0.05\\ SNR\_w = 0.20\end{tabular}}} & PCR & 3.00 (0.00) & \textbf{1.0 (0.00)} & \textbf{0.392 (0.005)} & \textbf{0.063 (0.005)} & \textbf{0.987 (0.002)} & \textbf{0.719 (0.013)} & 2.3 (0.31) \\
\multicolumn{1}{c|}{} & \multicolumn{1}{c|}{} & \multicolumn{1}{c|}{} & PLS & 2.50 (0.72) & 0.2 (0.13) & 1.204 (0.055) & 0.419 (0.078) & 0.708 (0.091) & 0.729 (0.014) & 2.1 (0.04) \\
\multicolumn{1}{c|}{} & \multicolumn{1}{c|}{} & \multicolumn{1}{c|}{} & SPC & 4.40 (0.79) & 0.7 (0.15) & 0.591 (0.021) & 0.187 (0.022) & 0.926 (0.024) & 0.722 (0.015) & 9.0 (0.17) \\
\multicolumn{1}{c|}{} & \multicolumn{1}{c|}{} & \multicolumn{1}{c|}{} & SOLAR & 3.00 (0.00) & \textbf{1.0 (0.00)} & \textbf{0.392 (0.005)} & 0.064 (0.005) & \textbf{0.987 (0.002)} & \textbf{0.719 (0.013)} & 23.5 (1.03) \\ \cline{2-11} 
\multicolumn{1}{c|}{} & \multicolumn{1}{c|}{\multirow{12}{*}{5}} & \multicolumn{1}{c|}{\multirow{4}{*}{\begin{tabular}[c]{@{}c@{}}SNR\_X = 0.20\\ SNR\_w = 0.05\end{tabular}}} & PCR & 1.60 (0.27) & 0.0 (0.00) & \textbf{0.364 (0.029)} & 0.301 (0.018) & 0.462 (0.107) & 1.643 (0.023) & 1.5 (0.12) \\
\multicolumn{1}{c|}{} & \multicolumn{1}{c|}{} & \multicolumn{1}{c|}{} & PLS & 4.20 (0.44) & 0.1 (0.10) & 1.695 (0.040) & 1.583 (0.025) & 0.195 (0.014) & 1.639 (0.023) & 2.7 (0.18) \\
\multicolumn{1}{c|}{} & \multicolumn{1}{c|}{} & \multicolumn{1}{c|}{} & SPC & 4.20 (1.07) & 0.1 (0.10) & 0.818 (0.076) & 0.445 (0.099) & 0.782 (0.065) & 1.636 (0.028) & 7.3 (0.11) \\
\multicolumn{1}{c|}{} & \multicolumn{1}{c|}{} & \multicolumn{1}{c|}{} & SOLAR & 5.00 (0.00) & \textbf{1.0 (0.00)} & 0.503 (0.011) & \textbf{0.161 (0.011)} & \textbf{0.909 (0.016)} & \textbf{1.617 (0.027)} & 26.2 (0.34) \\ \cline{3-11} 
\multicolumn{1}{c|}{} & \multicolumn{1}{c|}{} & \multicolumn{1}{c|}{\multirow{4}{*}{\begin{tabular}[c]{@{}c@{}}SNR\_X = 0.01\\ SNR\_w = 0.50\end{tabular}}} & PCR & 5.00 (0.00) & \textbf{1.0 (0.00)} & 1.071 (0.009) & \textbf{0.131 (0.002)} & \textbf{0.931 (0.003)} & 0.562 (0.008) & 1.5 (0.13) \\
\multicolumn{1}{c|}{} & \multicolumn{1}{c|}{} & \multicolumn{1}{c|}{} & PLS & 1.00 (0.00) & 0.0 (0.00) & 1.173 (0.009) & 0.385 (0.009) & 0.756 (0.010) & \textbf{0.546 (0.007)} & 2.3 (0.19) \\
\multicolumn{1}{c|}{} & \multicolumn{1}{c|}{} & \multicolumn{1}{c|}{} & SPC & 3.60 (0.76) & 0.0 (0.00) & 1.649 (0.157) & 0.342 (0.011) & 0.783 (0.018) & 0.557 (0.007) & 9.4 (0.16) \\
\multicolumn{1}{c|}{} & \multicolumn{1}{c|}{} & \multicolumn{1}{c|}{} & SOLAR & 1.50 (0.17) & 0.0 (0.00) & \textbf{0.554 (0.025)} & 0.304 (0.006) & 0.500 (0.026) & 0.610 (0.007) & 13.3 (0.80) \\ \cline{3-11} 
\multicolumn{1}{c|}{} & \multicolumn{1}{c|}{} & \multicolumn{1}{c|}{\multirow{4}{*}{\begin{tabular}[c]{@{}c@{}}SNR\_X = 0.05\\ SNR\_w = 0.20\end{tabular}}} & PCR & 4.30 (0.37) & 0.7 (0.15) & \textbf{0.568 (0.020)} & 0.116 (0.014) & 0.942 (0.015) & 0.817 (0.014) & 1.7 (0.20) \\
\multicolumn{1}{c|}{} & \multicolumn{1}{c|}{} & \multicolumn{1}{c|}{} & PLS & 2.90 (0.71) & 0.2 (0.13) & 1.471 (0.115) & 0.588 (0.070) & 0.622 (0.085) & 0.822 (0.014) & 2.4 (0.16) \\
\multicolumn{1}{c|}{} & \multicolumn{1}{c|}{} & \multicolumn{1}{c|}{} & SPC & 5.40 (0.86) & 0.4 (0.16) & 1.133 (0.097) & 0.310 (0.024) & 0.877 (0.019) & 0.822 (0.013) & 8.4 (0.28) \\
\multicolumn{1}{c|}{} & \multicolumn{1}{c|}{} & \multicolumn{1}{c|}{} & SOLAR & 5.00 (0.00) & \textbf{1.0 (0.00)} & 0.578 (0.010) & \textbf{0.093 (0.005)} & \textbf{0.970 (0.004)} & \textbf{0.814 (0.013)} & 27.2 (0.47) \\ \hline
\multicolumn{1}{c|}{\multirow{24}{*}{$(1000, 10^4)$}} & \multicolumn{1}{c|}{\multirow{12}{*}{3}} & \multicolumn{1}{c|}{\multirow{4}{*}{\begin{tabular}[c]{@{}c@{}}SNR\_X = 0.20\\ SNR\_w = 0.05\end{tabular}}} & PCR & 3.00 (0.00) & \textbf{1.0 (0.00)} & \textbf{0.369 (0.005)} & 0.073 (0.008) & \textbf{0.964 (0.009)} & 1.005 (0.016) & 5.5 (0.79) \\
\multicolumn{1}{c|}{} & \multicolumn{1}{c|}{} & \multicolumn{1}{c|}{} & PLS & 1.00 (0.00) & 0.0 (0.00) & 0.955 (0.065) & 0.141 (0.008) & 0.932 (0.015) & 1.011 (0.015) & 4.5 (0.30) \\
\multicolumn{1}{c|}{} & \multicolumn{1}{c|}{} & \multicolumn{1}{c|}{} & SPC & 7.50 (0.78) & 0.1 (0.10) & 0.481 (0.025) & 0.180 (0.020) & 0.869 (0.032) & 1.013 (0.014) & 21.9 (0.95) \\
\multicolumn{1}{c|}{} & \multicolumn{1}{c|}{} & \multicolumn{1}{c|}{} & SOLAR & 3.00 (0.00) & \textbf{1.0 (0.00)} & \textbf{0.369 (0.005)} & \textbf{0.068 (0.009)} & \textbf{0.964 (0.009)} & \textbf{1.004 (0.016)} & 45.8 (3.96) \\ \cline{3-11} 
\multicolumn{1}{c|}{} & \multicolumn{1}{c|}{} & \multicolumn{1}{c|}{\multirow{4}{*}{\begin{tabular}[c]{@{}c@{}}SNR\_X = 0.01\\ SNR\_w = 0.50\end{tabular}}} & PCR & 3.00 (0.00) & \textbf{1.0 (0.00)} & 0.574 (0.003) & \textbf{0.055 (0.001)} & \textbf{0.973 (0.001)} & 0.326 (0.005) & 5.5 (0.65) \\
\multicolumn{1}{c|}{} & \multicolumn{1}{c|}{} & \multicolumn{1}{c|}{} & PLS & 1.00 (0.00) & 0.0 (0.00) & 1.036 (0.017) & 0.157 (0.003) & 0.899 (0.005) & 0.326 (0.005) & 4.8 (0.25) \\
\multicolumn{1}{c|}{} & \multicolumn{1}{c|}{} & \multicolumn{1}{c|}{} & SPC & 3.80 (0.49) & 0.6 (0.16) & 0.981 (0.029) & 0.126 (0.004) & 0.924 (0.005) & 0.326 (0.005) & 22.1 (1.09) \\
\multicolumn{1}{c|}{} & \multicolumn{1}{c|}{} & \multicolumn{1}{c|}{} & SOLAR & 3.00 (0.00) & \textbf{1.0 (0.00)} & \textbf{0.573 (0.003)} & \textbf{0.055 (0.001)} & \textbf{0.973 (0.001)} & \textbf{0.325 (0.005)} & 48.0 (2.26) \\ \cline{3-11} 
\multicolumn{1}{c|}{} & \multicolumn{1}{c|}{} & \multicolumn{1}{c|}{\multirow{4}{*}{\begin{tabular}[c]{@{}c@{}}SNR\_X = 0.05\\ SNR\_w = 0.20\end{tabular}}} & PCR & 3.00 (0.00) & \textbf{1.0 (0.00)} & \textbf{0.400 (0.004)} & \textbf{0.042 (0.004)} & \textbf{0.987 (0.002)} & \textbf{0.502 (0.008)} & 5.6 (0.76) \\
\multicolumn{1}{c|}{} & \multicolumn{1}{c|}{} & \multicolumn{1}{c|}{} & PLS & 1.00 (0.00) & 0.0 (0.00) & 0.965 (0.032) & 0.115 (0.005) & 0.952 (0.007) & 0.509 (0.007) & 5.1 (0.53) \\
\multicolumn{1}{c|}{} & \multicolumn{1}{c|}{} & \multicolumn{1}{c|}{} & SPC & 5.80 (0.80) & 0.3 (0.15) & 0.573 (0.046) & 0.095 (0.007) & 0.963 (0.008) & 0.505 (0.007) & 21.6 (0.95) \\
\multicolumn{1}{c|}{} & \multicolumn{1}{c|}{} & \multicolumn{1}{c|}{} & SOLAR & 3.00 (0.00) & \textbf{1.0 (0.00)} & \textbf{0.400 (0.004)} & \textbf{0.042 (0.004)} & \textbf{0.987 (0.002)} & \textbf{0.502 (0.008)} & 44.5 (1.96) \\ \cline{2-11} 
\multicolumn{1}{c|}{} & \multicolumn{1}{c|}{\multirow{12}{*}{5}} & \multicolumn{1}{c|}{\multirow{4}{*}{\begin{tabular}[c]{@{}c@{}}SNR\_X = 0.20\\ SNR\_w = 0.05\end{tabular}}} & PCR & 3.30 (0.42) & 0.3 (0.15) & \textbf{0.414 (0.023)} & 0.131 (0.016) & 0.834 (0.047) & 1.107 (0.025) & 4.1 (0.36) \\
\multicolumn{1}{c|}{} & \multicolumn{1}{c|}{} & \multicolumn{1}{c|}{} & PLS & 2.20 (0.66) & 0.0 (0.00) & 1.235 (0.113) & 0.459 (0.139) & 0.707 (0.113) & 1.121 (0.027) & 4.3 (0.19) \\
\multicolumn{1}{c|}{} & \multicolumn{1}{c|}{} & \multicolumn{1}{c|}{} & SPC & 5.70 (1.11) & 0.2 (0.13) & 0.731 (0.095) & 0.176 (0.018) & 0.890 (0.019) & 1.112 (0.024) & 17.0 (0.66) \\
\multicolumn{1}{c|}{} & \multicolumn{1}{c|}{} & \multicolumn{1}{c|}{} & SOLAR & 5.00 (0.00) & \textbf{1.0 (0.00)} & 0.489 (0.005) & \textbf{0.084 (0.007)} & \textbf{0.947 (0.010)} & \textbf{1.103 (0.024)} & 51.2 (1.02) \\ \cline{3-11} 
\multicolumn{1}{c|}{} & \multicolumn{1}{c|}{} & \multicolumn{1}{c|}{\multirow{4}{*}{\begin{tabular}[c]{@{}c@{}}SNR\_X = 0.01\\ SNR\_w = 0.50\end{tabular}}} & PCR & 5.00 (0.00) & \textbf{1.0 (0.00)} & 0.929 (0.004) & \textbf{0.075 (0.001)} & \textbf{0.956 (0.001)} & 0.367 (0.008) & 4.1 (0.42) \\
\multicolumn{1}{c|}{} & \multicolumn{1}{c|}{} & \multicolumn{1}{c|}{} & PLS & 1.00 (0.00) & 0.0 (0.00) & 1.130 (0.012) & 0.209 (0.004) & 0.854 (0.005) & \textbf{0.362 (0.008)} & 4.4 (0.28) \\
\multicolumn{1}{c|}{} & \multicolumn{1}{c|}{} & \multicolumn{1}{c|}{} & SPC & 4.80 (0.44) & 0.4 (0.16) & 1.533 (0.057) & 0.169 (0.002) & 0.886 (0.004) & 0.363 (0.009) & 23.6 (2.02) \\
\multicolumn{1}{c|}{} & \multicolumn{1}{c|}{} & \multicolumn{1}{c|}{} & SOLAR & 4.90 (0.10) & 0.9 (0.10) & \textbf{0.919 (0.010)} & 0.079 (0.005) & 0.948 (0.009) & 0.367 (0.008) & 52.5 (1.35) \\ \cline{3-11} 
\multicolumn{1}{c|}{} & \multicolumn{1}{c|}{} & \multicolumn{1}{c|}{\multirow{4}{*}{\begin{tabular}[c]{@{}c@{}}SNR\_X = 0.05\\ SNR\_w = 0.20\end{tabular}}} & PCR & 5.00 (0.00) & \textbf{1.0 (0.00)} & \textbf{0.557 (0.004)} & \textbf{0.050 (0.003)} & \textbf{0.982 (0.002)} & \textbf{0.553 (0.012)} & 4.4 (0.64) \\
\multicolumn{1}{c|}{} & \multicolumn{1}{c|}{} & \multicolumn{1}{c|}{} & PLS & 1.80 (0.61) & 0.0 (0.00) & 1.204 (0.106) & 0.235 (0.053) & 0.832 (0.075) & 0.563 (0.013) & 4.7 (0.20) \\
\multicolumn{1}{c|}{} & \multicolumn{1}{c|}{} & \multicolumn{1}{c|}{} & SPC & 5.70 (0.45) & 0.4 (0.16) & 0.795 (0.006) & 0.118 (0.007) & 0.953 (0.006) & 0.556 (0.013) & 17.5 (0.60) \\
\multicolumn{1}{c|}{} & \multicolumn{1}{c|}{} & \multicolumn{1}{c|}{} & SOLAR & 5.00 (0.00) & \textbf{1.0 (0.00)} & \textbf{0.557 (0.004)} & \textbf{0.050 (0.003)} & \textbf{0.982 (0.002)} & \textbf{0.553 (0.012)} & 52.9 (1.40) \\ \hline
\end{tabular}}
\caption{
Representative moderate- and high-dimensional simulation results comparing PCR, PLS, SPC, and SOLAR across varying signal-to-noise regimes and latent ranks. Reported values are means with standard errors across 10 independent replications. Rank recovery proportion denotes the proportion of replications satisfying $\widehat q=q_{\mathrm{true}}$. The $H$-projector error measures recovery of the latent sample-level subspace. Supervised signal RMSE and correlation evaluate recovery of the latent supervised signal $H_{\mathrm{true}}\beta_{\mathrm{true}}$, while test RMSE evaluates out-of-sample prediction accuracy for the response $w$.
}
\label{tab:sim_supp}
\end{table}

\subsection{Preprocessing}

For all methods in the moderate- and high-dimensional comparison study, training predictors were column-centered using training-set means only. Test predictors were centered using the corresponding training-set means. The response was centered using the training-set response mean. No variance standardization of predictors was applied. Thus, all methods were fit to centered training data, and test predictions were transformed back to the original response scale by adding the training response mean.

For ultra-high-dimensional SOLAR simulations, the full generated matrix \(X\) and response \(w\) were centered prior to fitting, since no train-test split was used.

\subsection{Methods compared}

\paragraph{Principal component regression.}
PCR was fit by computing a truncated singular value decomposition of the centered training matrix,
\[
X_{\mathrm{train},c}\approx U_qS_qV_q^\top,
\]
and regressing the centered training response on the first \(q\) PC score vectors \(U_qS_q\). Candidate ranks \(q\in\{1,\ldots,10\}\) were considered, subject to the constraint \(q<n_{\mathrm{train}}\). The number of components was selected by minimizing the training BIC
\[
\mathrm{BIC}(q)
=
n_{\mathrm{train}}\log\{\mathrm{RSS}(q)/n_{\mathrm{train}}\}
+
q\log(n_{\mathrm{train}}),
\]
where \(\mathrm{RSS}(q)\) is the residual sum of squares from regressing the centered training response on the first \(q\) PC scores. Final PCR predictions on the test set were obtained by projecting centered test predictors onto the selected loading vectors and applying the fitted regression coefficients.

\paragraph{Partial least squares.}
PLS was implemented using MATLAB's \texttt{plsregress} function. Candidate component numbers \(q\in\{1,\ldots,10\}\) were considered, again subject to \(q<n_{\mathrm{train}}\). The number of PLS components was selected by five-fold cross-validation within the training set. Within each fold, the fold-specific training predictors and response were centered, PLS was fit on the fold-specific training portion, and prediction error was evaluated on the validation fold. The selected rank minimized the cross-validated RMSE. After rank selection, PLS was refit on the full training set and evaluated on the held-out test set.

For latent subspace evaluation, PLS scores and \(X\)-loadings were not treated as an SVD factorization. Instead, the estimated sample and feature subspaces were represented by orthonormal bases obtained from the fitted PLS score matrix and loading matrix, respectively. This allowed rotation-invariant comparison with the true latent subspaces.

\paragraph{Supervised principal components.}
SPC followed the supervised principal components framework of \citet{bair2006prediction}. For each training set, variables were screened using the absolute marginal association score
\[
|X_{\mathrm{train},c}^{\top}w_{\mathrm{train},c}|.
\]
The candidate screened-variable counts were
\[
p_S\in
\mathrm{unique}\{\mathrm{round}(p\times a):a\in(0.01,0.025,0.05,0.10,0.20)\}.
\]
Five-fold cross-validation jointly selected the screening size \(p_S\) and the number of principal components \(q\in\{1,\ldots,10\}\). In each fold, screening was performed using only the fold-specific training data, PCA was applied to the screened training predictors, and validation error was computed from regression on the resulting supervised PC scores. The final SPC model was fit on the full training set using the selected screening size and rank, and test prediction was performed using the corresponding screened test predictors.

\paragraph{SOLAR.}
SOLAR was fit using the penalized MAP objective described in Section~3 of the main manuscript. The candidate rank range was \(q\in\{1,\ldots,10\}\), with initialization at \(q_{\mathrm{init}}=10\). The dimension-adaptive penalty used
\[
\kappa(n,p)=
c_{\kappa}
\left\{
\frac{(\sqrt n+\sqrt p)^2}{n+p}
\right\}^{\zeta},
\]
with \(c_{\kappa}=0.1\) and \(\zeta=2/3\). For moderate- and high-dimensional split-based simulations, \(n\) in this expression was replaced by \(n_{\mathrm{train}}\). The prior variances were fixed at \(\rho^2=0.5^2\) and \(g^2=25\). The residual variance \(\sigma^2\) was estimated from a rank-2 baseline residual approximation, and \(\tau^2\) was initialized by the empirical variance of the centered response.

The optimization used alternating closed-form Stiefel updates for \(V\), \(H\), \(\beta\), and \(D\), together with trans-dimensional rank proposals. The simulated annealing temperature was
\[
T_t=\max\{T_{\min},T_0t^{-0.7}\},
\]
with \(T_0=1\) and \(T_{\min}=10^{-2}\). At each iteration, a rank move was proposed with probability 0.5. Proposed moves changed the rank by one, subject to the bounds \(1\le q\le 10\), and were accepted using the Metropolis rule based on the change in penalized objective and current temperature. The best MAP state observed over the run was retained as the final estimator.

For computational efficiency, SOLAR used a Gram-based SVD initialization and updates. Specifically, for centered \(X\), the algorithm formed \(G=XX^\top\), computed the leading eigenvectors of \(G\), and recovered the corresponding right singular vectors as \(X^\top U S^{-1}\). This avoids forming the \(p\times p\) covariance matrix and is essential for the ultra-high-dimensional settings.

For split-based simulations, SOLAR test predictions were obtained by projecting centered test predictors onto the estimated loading space:
\[
\widehat H_{\mathrm{test}}
=
X_{\mathrm{test},c}\widehat V\widehat D^{-1},
\]
with small diagonal entries of \(\widehat D\) truncated away from zero for numerical stability. The test prediction was then
\[
\widehat w_{\mathrm{test}}
=
\widehat H_{\mathrm{test}}\widehat\beta
+
\bar w_{\mathrm{train}}.
\]

\subsection{SOLAR stopping criteria}

The trans-dimensional simulated annealing stage of SOLAR was terminated using a joint stability criterion designed to detect practical convergence of the penalized objective and stabilization of the selected latent dimension. Let \(W_{\mathrm{stop}}\) denote a sliding monitoring window over recent iterations. For moderate- and high-dimensional simulations, we used $W_{\mathrm{stop}}=1000,$ while ultra-high-dimensional simulations used $W_{\mathrm{stop}}=250.$ Termination required simultaneous satisfaction of the following conditions:

\begin{enumerate}
\item[(i)] \textbf{Temperature stabilization:}
the annealing temperature had effectively reached the minimum temperature floor, $T_t \le (1+\eta_T)T_{\min},$ with \(\eta_T=0.10, T_{\min}=10^{-2}\).

\item[(ii)] \textbf{Rank stabilization:}
the best selected rank remained unchanged throughout the monitoring window of $W_{\mathrm{stop}}$.

\item[(iii)] \textbf{Objective stabilization:}
the relative improvement in the best penalized objective value over the monitoring window satisfied
\[
\frac{
|J_t^{\mathrm{best}}
-
J_{t-W_{\mathrm{stop}}+1}^{\mathrm{best}}|
}{
1+|J_t^{\mathrm{best}}|
}
\le 10^{-6}.
\]

\item[(iv)] \textbf{Vanishing trans-dimensional exploration:}
the empirical acceptance rate of rank-move proposals within the monitoring window was at most \(0.01\).
\end{enumerate}

Collectively, these criteria indicate that the annealing dynamics have entered a stable low-temperature regime, the selected latent dimension has ceased fluctuating, the penalized objective has effectively plateaued, and trans-dimensional exploration has largely terminated. The final SOLAR estimator was taken as the parameter configuration achieving the largest penalized objective value encountered during the optimization trajectory.

\subsection{Evaluation metrics}

All reported summaries are means with standard errors over 10 independent replications. For each replication, the estimated rank is denoted \(\widehat q\), and exact rank recovery is
\[
I(\widehat q=q_0).
\]
The reported rank recovery proportion is the average of this indicator over replications. The absolute rank error is
\[
|\widehat q-q_0|.
\]

Because the latent factors are rotationally non-identifiable, subspace recovery was assessed using projector-based errors. Let
\[
q_c=\min(\widehat q,q_0).
\]
The \(H\)-projector error was computed as
\[
\left\|
H_{0,c}H_{0,c}^{\top}
-
\widehat H_c\widehat H_c^{\top}
\right\|_F,
\]
where \(H_{0,c}\) and \(\widehat H_c\) denote the first \(q_c\) columns of the true and estimated sample-level latent factor matrices. In split-based simulations, this metric was computed on the training sample. 

Low-rank signal reconstruction error was computed without explicitly forming large reconstructed matrices. For methods with an SVD-like representation \(\widehat H\widehat D\widehat V^\top\), the squared signal reconstruction error was evaluated as
\[
\|H_{0,c}D_{0,c}V_{0,c}^\top-\widehat H_c\widehat D_c\widehat V_c^\top\|_F^2
=
\|D_{0,c}\|_F^2+\|\widehat D_c\|_F^2
-
2\,\mathrm{tr}
\left\{
D_{0,c}(H_{0,c}^{\top}\widehat H_c)
\widehat D_c(\widehat V_c^\top V_{0,c})
\right\}.
\]
The relative \(X\)-signal reconstruction error divided the square root of this quantity by \(\|D_{0,c}\|_F\), and the entrywise \(X\)-signal RMSE divided the square root by \((n_{\mathrm{train}}p)^{1/2}\) in split-based simulations or by \((np)^{1/2}\) in no-split simulations. For PLS, which does not directly estimate a diagonal SVD-like loading matrix, the projected core matrix
\[
\widehat C=\widehat H^\top X_{\mathrm{train},c}\widehat V
\]
was used in place of \(\widehat D\).

Supervised signal recovery was assessed by comparing the estimated supervised latent signal with the true signal. In split-based simulations, the true supervised signal on the training sample was
\[
w_{0,\mathrm{train}}=H_{0,\mathrm{train}}\beta_0.
\]
The estimated supervised signal was the centered training fitted value from each method. We reported supervised signal RMSE,
\[
\frac{\|\widehat w_{\mathrm{signal}}-w_{0,\mathrm{train}}\|_2}{\sqrt{n_{\mathrm{train}}}},
\]
and supervised signal correlation,
\[
\mathrm{corr}(\widehat w_{\mathrm{signal}},w_{0,\mathrm{train}}).
\]
In ultra-high-dimensional no-split simulations, the same metrics were computed using the full sample.

Out-of-sample prediction metrics were computed only for moderate- and high-dimensional split-based simulations. The primary prediction metric was test RMSE,
\[
\left\{
\frac{1}{n_{\mathrm{test}}}
\sum_{i\in\mathcal I_{\mathrm{test}}}
(w_i-\widehat w_i)^2
\right\}^{1/2}.
\]
We also recorded test MAE, test correlation, and test \(R^2\), although only test RMSE is emphasized in the main text.

For methods where \(\widehat q=q_0\), beta recovery was additionally assessed after Procrustes alignment of the estimated and true \(H\)-spaces. Let
\[
H_{0,\mathrm{train}}^\top \widehat H = U\Sigma V^\top,
\qquad
R=UV^\top.
\]
The aligned coefficient was \(R^\top\widehat\beta\), and beta error was reported as
\[
\|R^\top\widehat\beta-\beta_0\|_2.
\]
This metric was recorded for completeness but not emphasized in the main paper because it is only directly comparable when the estimated and true ranks agree.

\subsection{Summary construction}

Each method produced one CSV file per replication and scenario, containing the estimated rank, recovery metrics, prediction metrics, and runtime. Scenario-level summaries were then computed by averaging each metric over the available replications. Reported uncertainty values are standard errors, computed as the empirical standard deviation divided by the square root of the number of nonmissing replications. Combined summary tables were created by merging the method-specific summaries for PCR, PLS, SPC, and SOLAR.

For moderate- and high-dimensional simulations, the summary tables include PCR, PLS, SPC, and SOLAR. For ultra-high-dimensional simulations, only SOLAR was summarized. Runtime was recorded using MATLAB's \texttt{tic}/\texttt{toc} around the model fitting and evaluation steps. Moderate- and high-dimensional runs used single-core execution on nodes with 256 GB RAM. Ultra-high-dimensional SOLAR runs used 494 GB RAM and 40-thread multithreaded execution.


\section{Technical Assumptions and Detailed Proofs}\label{sec:supp_theory}
This section provides the detailed assumptions and technical framework underlying the theoretical results stated in the main manuscript. Throughout, we consider the supervised factor model
\[
X = H_0D_0V_0^\top + E,
\qquad
w = H_0\beta_0 + \varepsilon,
\]
where \(H_0\in\mathrm{St}(n,q_0)\), \(V_0\in\mathrm{St}(p,q_0)\), \(D_0\in\mathbb{R}^{q_0\times q_0}\) is diagonal with positive entries, and \(\beta_0\in\mathbb{R}^{q_0}\). The noise matrices satisfy
\[
E_{ij}\overset{\mathrm{iid}}{\sim}N(0,\sigma^2),
\qquad
\varepsilon_i\overset{\mathrm{iid}}{\sim}N(0,\tau^2),
\]
with \(E\) and \(\varepsilon\) independent.

\subsection{Regularity Conditions}

\begin{enumerate}

\item[(A1)] \textbf{Spectral separation, signal strength, and canonical identifiability.}
The true rank \(q_0\) is fixed, and the nonzero diagonal entries of \(D_0\) satisfy
\[
d_{01}>d_{02}>\cdots>d_{0q_0}>0.
\]
Moreover, there exists \(\delta>0\) such that
\[
\frac{d_{0q_0}}
{\sigma(\sqrt n+\sqrt p)}
\ge 1+\delta
\]
eventually, and the relative eigengap is bounded away from zero:
\[
\min_{1\le k<q_0}
\frac{d_{0k}-d_{0,k+1}}{d_{0k}}
\ge c_D
\]
for some constant \(c_D>0\).

\item[(A2)] \textbf{Supervised signal strength and estimability.}
The supervised coefficient vector satisfies
\[
\|\beta_0\|_2\ge c_\beta>0.
\]
Under the Stiefel normalization \(H_0^\top H_0=I_{q_0}\), we additionally assume that the supervised noise variance satisfies
\[
\tau_n^2\to 0,
\]
or, equivalently, that the effective supervised signal-to-noise level increases sufficiently for \(\beta_0\) to be consistently estimable. This condition is needed because, with orthonormal columns of \(H_0\), fixed \(\tau^2\) would yield nonvanishing uncertainty in \(H_0^\top w\).

\item[(A3)] \textbf{Controlled dimensional growth.}
As \(n\to\infty\), the ambient dimension \(p=p_n\) may grow with \(n\), but satisfies
\[
\log p=o(n),
\]
and the true rank \(q_0\) remains fixed.

\item[(A4)] \textbf{Dimension-adaptive penalty scaling.}
The penalty multiplier satisfies
\[
\kappa(n,p)
=
c_{\kappa}
\left\{
\frac{(\sqrt n+\sqrt p)^2}{n+p}
\right\}^{\zeta},
\qquad
c_{\kappa}>0,\quad \zeta>0,
\]
and there exist constants \(0<c_{\min}\le c_{\max}<\infty\), independent of \(n\) and \(p\), such that
\[
c_{\min}
\le
\kappa(n,p)
\le
c_{\max}.
\]
The penalized objective uses the BIC-type penalty
\[
\kappa(n,p)\frac12\log(np)\,\mathrm{df}(q).
\]

\item[(A5)] \textbf{Bounded candidate rank space.}
The rank search is over
\[
1\le q\le q_{\max},
\]
where \(q_{\max}\) is fixed and \(q_0\le q_{\max}\).

\item[(A6)] \textbf{Signal detectability for rank selection.}
The weakest latent signal satisfies
\[
\frac{d_{0q_0}^2/\sigma^2}
{(n+p)\log(np)}
\to\infty.
\]
This condition ensures that the smallest true latent factor remains asymptotically distinguishable from noise under the BIC-type penalty.
\end{enumerate}

Assumption (A1) imposes the standard spectral separation needed for stable recovery of low-rank latent structure. The requirement that the weakest signal singular value remains separated from the Gaussian noise spectral edge ensures that the true factors are distinguishable from noise-driven singular directions, while the relative eigengap condition supports stable subspace recovery under perturbation \citep{davis1970rotation,wedin1972perturbation}. Distinct singular values further reduce the remaining nonidentifiability to simultaneous signed permutations, enabling a canonical representation.

Assumption (A2) prevents degeneracy of the supervised component. Because the columns of \(H_0\) are orthonormal under the Stiefel normalization, consistent recovery of \(\beta_0\) requires the effective supervised noise level to vanish asymptotically, or equivalently that the supervised signal-to-noise level increases sufficiently. This condition ensures that the response contains enough information to identify the supervised direction within the recovered latent space.

Assumption (A3) permits high-dimensional asymptotics while excluding exponentially growing ambient dimensions. This type of growth condition is standard in high-dimensional factor and low-rank matrix recovery settings \citep{bai2003inferential,johnstone2009consistency}, where concentration of noise terms and uniform control over candidate low-rank models are required. The fixed-rank condition reflects the low-dimensional latent structure targeted by SOLAR.

Assumption (A4) formalizes the dimension-adaptive penalty used in the proposed objective. In the simulations and applications, we use
\[
\kappa(n,p)
=
c_{\kappa}
\left\{
\frac{(\sqrt n+\sqrt p)^2}{n+p}
\right\}^{\zeta},
\qquad
c_{\kappa}>0,\quad \zeta>0.
\]
Since
\[
1
\le
\frac{(\sqrt n+\sqrt p)^2}{n+p}
=
1+\frac{2\sqrt{np}}{n+p}
\le
2,
\]
it follows that
\[
c_{\kappa}
\le
\kappa(n,p)
\le
2^{\zeta}c_{\kappa}.
\]
Thus, \(\kappa(n,p)\) is uniformly bounded away from zero and infinity for all \(n,p\). Consequently, the rank penalty retains the same asymptotic BIC order as the classical information criterion of \citet{schwarz1978estimating}, namely
\[
\frac12\log(np)\,\mathrm{df}(q),
\]
while allowing finite-sample dimension adaptation through a bounded multiplicative factor.

Assumption (A5) ensures that rank selection is performed over a fixed finite model class containing the true rank. This avoids diverging model complexity and permits the rank-selection argument to compare finitely many underfitted, correctly specified, and overfitted candidate models.

Assumption (A6) is an additional signal-detectability condition used only for the rank-selection consistency result. It plays a role analogous to beta-min conditions in high-dimensional variable-selection theory by ensuring that the weakest latent factor remains asymptotically distinguishable from noise after accounting for the BIC-type complexity penalty. Similar signal-strength requirements are common in consistent model-selection arguments for latent-factor and low-rank estimation problems. This condition is sufficient rather than minimal and is introduced primarily to establish asymptotic rank-selection consistency under the proposed BIC-type penalization framework.
\subsection{Proof of Proposition~1}
\begin{proof}
Let
\[
M_0 = H_0D_0V_0^\top
\]
denote the noiseless low-rank mean matrix. Under (A1), the nonzero diagonal entries of \(D_0\) are strictly positive, distinct, and separated. Hence \(M_0\) has rank \(q_0\), and its nonzero singular values are precisely the diagonal entries of \(D_0\), up to ordering. By the uniqueness of the singular value decomposition when the nonzero singular values are distinct, the associated left and right singular vectors are identifiable up to simultaneous sign changes and permutations of the latent components; see, for example, standard singular value decomposition uniqueness results in \citet{stewart1990matrix,horn2012matrix}.

To make this explicit, suppose that another representation satisfies
\[
M_0 = H D V^\top,
\]
where \(H\in\mathrm{St}(n,q_0)\), \(V\in\mathrm{St}(p,q_0)\), and \(D\) is diagonal with strictly positive and distinct diagonal entries. Since both decompositions represent the same rank-\(q_0\) matrix with distinct nonzero singular values, there exists a signed permutation matrix \(P\) such that
\[
H = H_0P,\qquad
V = V_0P,\qquad
D = P^\top D_0P.
\]
Here \(P\) has exactly one nonzero entry in each row and column, each equal to \(+1\) or \(-1\). Because \(P\) is a signed permutation matrix, \(P^\top D_0P\) remains diagonal and the representation remains within the SOLAR parameter space.

Now consider the supervised component. The noiseless supervised signal is
\[
m_0 = H_0\beta_0.
\]
Under the alternative representation \(H=H_0P\), the same signal can be written as \(m_0=H\beta\) if and only if
\[
H_0\beta_0 = H_0P\beta.
\]
Premultiplying by \(H_0^\top\) and using \(H_0^\top H_0=I_{q_0}\) gives
\[
\beta_0=P\beta,
\qquad\text{or equivalently}\qquad
\beta=P^\top\beta_0.
\]
Thus the full supervised factor tuple is identifiable up to the simultaneous signed permutation
\[
(H_0,V_0,D_0,\beta_0)
\equiv
(H_0P,\;V_0P,\;P^\top D_0P,\;P^\top\beta_0).
\]

It remains to verify that this ambiguity does not affect fitted quantities or the objective. For any signed permutation matrix \(P\), define
\[
\widetilde H=HP,\qquad
\widetilde V=VP,\qquad
\widetilde D=P^\top DP,\qquad
\widetilde\beta=P^\top\beta.
\]
Then
\[
\widetilde H\widetilde D\widetilde V^\top
=
(HP)(P^\top DP)(VP)^\top
=
HDV^\top,
\]
and
\[
\widetilde H\widetilde\beta
=
(HP)(P^\top\beta)
=
H\beta.
\]
Therefore the fitted low-rank mean matrix and fitted supervised signal are unchanged by the simultaneous signed permutation.

The objective function in Section~3 depends on \((H,V,D,\beta)\) through the residual norms
\[
\|X-HDV^\top\|_F^2,
\qquad
\|w-H\beta\|_2^2,
\]
and through the quadratic penalties on \(d=\mathrm{diag}(D)\) and \(\beta\). The two residual norms are unchanged by the identities above. Moreover, signed permutations preserve Euclidean norms, so
\[
\|\mathrm{diag}(\widetilde D)\|_2=\|\mathrm{diag}(D)\|_2,
\qquad
\|\widetilde\beta\|_2=\|\beta\|_2.
\]
Hence the likelihood, prior penalties, and penalized objective are all invariant under the remaining signed-permutation ambiguity.

Finally, the canonical orientation step used in SOLAR applies a deterministic convention to this remaining ambiguity. In particular, after the basis is re-diagonalized, the latent components can be ordered by their diagonal loading magnitudes and their signs fixed by a deterministic supervised-direction convention whenever the supervised direction is nondegenerate under (A2). This selects a single representative from each signed-permutation equivalence class. Since this operation only reorders and changes signs of corresponding columns of \(H\) and \(V\), while applying the same transformation to \(D\) and \(\beta\), it does not alter \(HDV^\top\), \(H\beta\), or the value of the objective function. This proves the proposition.
\end{proof}
\subsection{Proof of Theorem~1}

Under assumptions (A1)--(A5), the canonical fixed-rank population objective is uniquely maximized at the true parameter \(\theta_0\), up to the signed-permutation ambiguity resolved by Proposition~1. Consequently, the population objective is uniformly separated from its maximum outside any sufficiently small neighborhood of the canonical representative of \(\theta_0\).
\begin{proof}
Let \(q=q_0\) be fixed and write
\[
\theta=(H,V,D,\beta),\qquad \theta_0=(H_0,V_0,D_0,\beta_0).
\]
Define the fitted low-rank mean and supervised signal by
\[
M(\theta)=HDV^\top,\qquad m(\theta)=H\beta,
\]
with \(M_0=M(\theta_0)=H_0D_0V_0^\top\) and \(m_0=m(\theta_0)=H_0\beta_0\). Since \(q_0\) is fixed and the estimator is represented under the canonical orientation of Proposition~1, the only remaining signed-permutation ambiguity is removed.

We first establish consistency of the fitted mean objects. Using
\[
X=M_0+E,\qquad w=m_0+\varepsilon,
\]
the matrix residual difference satisfies
\[
\|X-M(\theta)\|_F^2-\|X-M_0\|_F^2
=
\|M(\theta)-M_0\|_F^2
-
2\langle E,M(\theta)-M_0\rangle.
\]
Similarly,
\[
\|w-m(\theta)\|_2^2-\|w-m_0\|_2^2
=
\|m(\theta)-m_0\|_2^2
-
2\varepsilon^\top\{m(\theta)-m_0\}.
\]
Therefore, after including the fixed-rank prior contributions,
\[
\ell(\theta)-\ell(\theta_0)
=
-\frac{1}{2\sigma^2}\|M(\theta)-M_0\|_F^2
+
\frac{1}{\sigma^2}\langle E,M(\theta)-M_0\rangle
\]
\[
\hspace{1.2cm}
-\frac{1}{2\tau_n^2}\|m(\theta)-m_0\|_2^2
+
\frac{1}{\tau_n^2}\varepsilon^\top\{m(\theta)-m_0\}
+
R_{\mathrm{prior}}(\theta,\theta_0),
\]
where
\[
R_{\mathrm{prior}}(\theta,\theta_0)
=
-\frac{1}{2\rho^2}\{\|d\|_2^2-\|d_0\|_2^2\}
-\frac{1}{2g^2}\{\|\beta\|_2^2-\|\beta_0\|_2^2\}.
\]
Under the regularity conditions, the deterministic likelihood separation dominates the stochastic fluctuation terms outside any fixed neighborhood of the true fitted objects. This follows from Gaussian concentration and standard compactness/argmax arguments for constrained \(M\)-estimation \citep{vanDerVaart1998asymptotic}.

More explicitly, for any \(\eta>0\), define
\[
\mathcal N_\eta^c
=
\left\{
\theta:
\frac{\|M(\theta)-M_0\|_F}{\|D_0\|_F}
+
\frac{\|m(\theta)-m_0\|_2}{\|\beta_0\|_2}
\ge \eta
\right\}.
\]
By (A1), the nonzero singular values of \(M_0\) lie above the high-dimensional noise spectral edge, and by (A2), the supervised component is estimable under the Stiefel normalization. Hence the deterministic part of the objective is uniformly separated from its maximum value at \(\theta_0\) on \(\mathcal N_\eta^c\). The stochastic terms are controlled uniformly over the fixed-rank parameter space using Gaussian spectral-norm bounds and standard concentration arguments; since \(q_0\) is fixed and the signal singular values are separated from the high-dimensional noise edge in (A1), these fluctuations are asymptotically dominated by the deterministic separation. Consequently,
\[
\Pr\left[
\sup_{\theta\in\mathcal N_\eta^c}
\{\ell(\theta)-\ell(\theta_0)\}<0
\right]\to1.
\]
Since \(\hat\theta=(\hat H,\hat V,\hat D,\hat\beta)\) maximizes the fixed-rank objective, it follows that
\[
\frac{\|M(\hat\theta)-M_0\|_F}{\|D_0\|_F}\xrightarrow{p}0,
\qquad
\|m(\hat\theta)-m_0\|_2\xrightarrow{p}0.
\]

We now translate fitted-object consistency into componentwise recovery. By (A1), \(D_0\) has positive, distinct, and spectrally separated diagonal entries. Standard singular subspace perturbation results of Davis--Kahan and Wedin imply that relative perturbation of \(M_0\) yields convergence of the associated left and right singular subspaces \citep{davis1970rotation,wedin1972perturbation,stewart1990matrix}. Hence,
\[
\hat H\hat H^\top \xrightarrow{p} H_0H_0^\top,
\qquad
\hat V\hat V^\top \xrightarrow{p} V_0V_0^\top.
\]
The same perturbation argument, together with the relative separation of the singular values, gives convergence of the ordered singular strengths in relative Frobenius norm:
\[
\frac{\|\hat D-D_0\|_F}{\|D_0\|_F}\xrightarrow{p}0.
\]

It remains to show recovery of \(\beta_0\). Since
\[
m(\hat\theta)=\hat H\hat\beta
\qquad\text{and}\qquad
m_0=H_0\beta_0,
\]
we have
\[
\|\hat H\hat\beta-H_0\beta_0\|_2\xrightarrow{p}0.
\]
The canonical orientation removes the signed-permutation ambiguity, so subspace convergence implies
\[
\hat H^\top H_0\xrightarrow{p}I_{q_0}.
\]
Premultiplying the supervised-signal error by \(\hat H^\top\), and using \(\hat H^\top\hat H=I_{q_0}\), gives
\[
\hat\beta-\hat H^\top H_0\beta_0=o_p(1).
\]
Therefore,
\[
\hat\beta
=
\hat H^\top H_0\beta_0+o_p(1)
=
\beta_0+o_p(1),
\]
which proves
\[
\|\hat\beta-\beta_0\|_2\xrightarrow{p}0.
\]

Combining these results establishes the asserted fixed-rank recovery. The consistency of the low-rank signal follows from
\[
\frac{\|\hat H\hat D\hat V^\top-H_0D_0V_0^\top\|_F}{\|D_0\|_F}\xrightarrow{p}0,
\]
and the consistency of the supervised signal follows from
\[
\|\hat H\hat\beta-H_0\beta_0\|_2\xrightarrow{p}0.
\]
\end{proof}

\subsection{Proof of Theorem~2}

\begin{proof}
For each candidate rank \(q\), define the maximized penalized objective
\[
\widehat J_q
=
\sup_{\theta_q}
\left\{
\ell(\theta_q;q)
-
\kappa(n,p)\frac12\log(np)\,\mathrm{df}(q)
\right\},
\]
where \(\theta_q=(H,V,D,\beta)\) ranges over the constrained rank-\(q\) parameter space. The selected rank satisfies
\[
\hat q=\arg\max_{1\le q\le q_{\max}}\widehat J_q.
\]
Since \(q_{\max}\) is fixed and \(q_0\le q_{\max}\) by (A5), it is enough to show that, with probability tending to one,
\[
\widehat J_q < \widehat J_{q_0}
\qquad
\text{for every }q\ne q_0.
\]
We consider underfitted and overfitted ranks separately.

First consider \(q<q_0\). Any rank-\(q\) matrix \(HDV^\top\) omits at least one nonzero latent factor from the true rank-\(q_0\) signal \(M_0=H_0D_0V_0^\top\). By the Eckart--Young--Mirsky theorem \citep{eckart1936approximation,mirsky1960symmetric},
\[
\inf_{\mathrm{rank}(M)\le q}
\|M-M_0\|_F^2
=
\sum_{k=q+1}^{q_0} d_{0k}^2
\ge d_{0q_0}^2.
\]
Thus any underfitted rank-\(q\) model incurs a deterministic matrix-fit loss of order at least \(d_{0q_0}^2/(2\sigma^2)\). Gaussian concentration and the growth conditions in (A1)--(A3) imply that, uniformly over the finite collection \(q<q_0\),
\[
\sup_{q<q_0}
\left[
\sup_{\theta_q}\ell(\theta_q;q)
-
\sup_{\theta_{q_0}}\ell(\theta_{q_0};q_0)
\right]
\le
-
c\frac{d_{0q_0}^2}{\sigma^2}
+
O_p(n+p),
\]
for some constant \(c>0\). The penalty favors underfitted models because \(\mathrm{df}(q)<\mathrm{df}(q_0)\), and the maximal possible penalty advantage is
\[
\kappa(n,p)\frac12\log(np)
\{\mathrm{df}(q_0)-\mathrm{df}(q)\}
=
O\{(n+p)\log(np)\},
\]
since \(q_{\max}\) is fixed and \(\mathrm{df}(q)=q(n+p-2q)+2q\). By the BIC-detectability condition in (A6),
\[
\frac{d_{0q_0}^2/\sigma^2}{(n+p)\log(np)}
\to\infty.
\]
Hence the likelihood loss from omitting a true factor dominates both the stochastic fluctuation and the penalty reduction. Therefore,
\[
\Pr\left(
\max_{q<q_0}\widehat J_q
<
\widehat J_{q_0}
\right)\to1.
\]

Now consider \(q>q_0\). The rank-\(q\) model contains the true rank-\(q_0\) model as a submodel by allowing the additional latent strengths and supervised coefficients to be zero. Hence overfitting can improve the fixed-rank log-posterior only by fitting noise directions. Standard Gaussian random matrix bounds give
\[
\|E\|_{\mathrm{op}}
=
O_p\{\sigma(\sqrt n+\sqrt p)\}
\]
\citep{davidson2001local}. Since \(q_{\max}\) is fixed, the total additional matrix-fit gain from adding finitely many spurious latent factors is therefore
\[
O_p\{(\sqrt n+\sqrt p)^2\}
=
O_p(n+p).
\]
The supervised part contributes at most \(O_p(1)\) additional improvement for each extra finite-dimensional component under (A2), and the fixed-rank prior terms remain controlled over the bounded rank range. Consequently, uniformly over \(q>q_0\),
\[
\sup_{\theta_q}\ell(\theta_q;q)
-
\sup_{\theta_{q_0}}\ell(\theta_{q_0};q_0)
=
O_p(n+p).
\]

The penalty increase from \(q_0\) to \(q>q_0\) is
\[
\kappa(n,p)\frac12\log(np)
\{\mathrm{df}(q)-\mathrm{df}(q_0)\}.
\]
Because \(q_{\max}\) is fixed and \(q>q_0\),
\[
\mathrm{df}(q)-\mathrm{df}(q_0)
=
(q-q_0)(n+p)-2(q^2-q_0^2)+2(q-q_0)
=
(q-q_0)(n+p)+O(1).
\]
By (A4), \(\kappa(n,p)\ge c_{\min}>0\). Therefore, for every \(q>q_0\),
\[
\kappa(n,p)\frac12\log(np)
\{\mathrm{df}(q)-\mathrm{df}(q_0)\}
\ge
c (n+p)\log(np)
\]
for some \(c>0\) and all sufficiently large \(n,p\). Since the overfitting gain is \(O_p(n+p)\), while the penalty increase is of order \((n+p)\log(np)\), the penalty dominates. Hence,
\[
\Pr\left(
\max_{q>q_0}\widehat J_q
<
\widehat J_{q_0}
\right)\to1.
\]

Combining the underfitting and overfitting comparisons yields
\[
\Pr\left(
\widehat J_{q_0}
>
\max_{q\ne q_0}\widehat J_q
\right)\to1.
\]
Therefore,
\[
\Pr(\hat q=q_0)\to1,
\]
which proves the theorem.
\end{proof}

\section{Additional results for the SOLAR case study}

Table~\ref{tab:OOS} summarizes an additional out-of-sample comparison for residualized Horvath DNAm age using a common 80\% training and 20\% test split across methods under a fixed latent dimension (\(q=5\)). Prediction accuracy was evaluated using root mean squared error (RMSE) and Pearson correlation between observed and predicted outcomes on the held-out test samples. In this real-data setting, PLS achieved the strongest predictive performance under the fixed low-rank specification, whereas SOLAR and PCR exhibited broadly comparable predictive behavior. This pattern is consistent with the primary objective of SOLAR, which emphasizes recovery of stable and interpretable supervised methylation structure associated with developmental biological-aging heterogeneity rather than purely prediction-optimized latent regression. In addition to competitive predictive performance, SOLAR further provides structured latent factor decomposition, CpG-level loading organization, developmental factor trajectory visualization, and biologically interpretable downstream enrichment characterization.

\begin{table}[t]
\centering
\renewcommand{\arraystretch}{1.1}
\resizebox{0.35\columnwidth}{!}{%
\begin{tabular}{cccc}
\hline
Method & Rank & RMSE & Corr. \\ \hline
PCR   & 5 & 0.69 & 0.42 \\
PLS   & 5 & 0.61 & 0.60 \\
SOLAR & 5 & 0.69 & 0.42 \\ \hline
\end{tabular}}
\caption{Out-of-sample prediction performance for residualized Horvath DNAm age under a fixed latent dimension (\(q=5\)).}
\label{tab:OOS}
\end{table}

\begin{figure}[!t]
\centering
\includegraphics[width=0.99\textwidth]{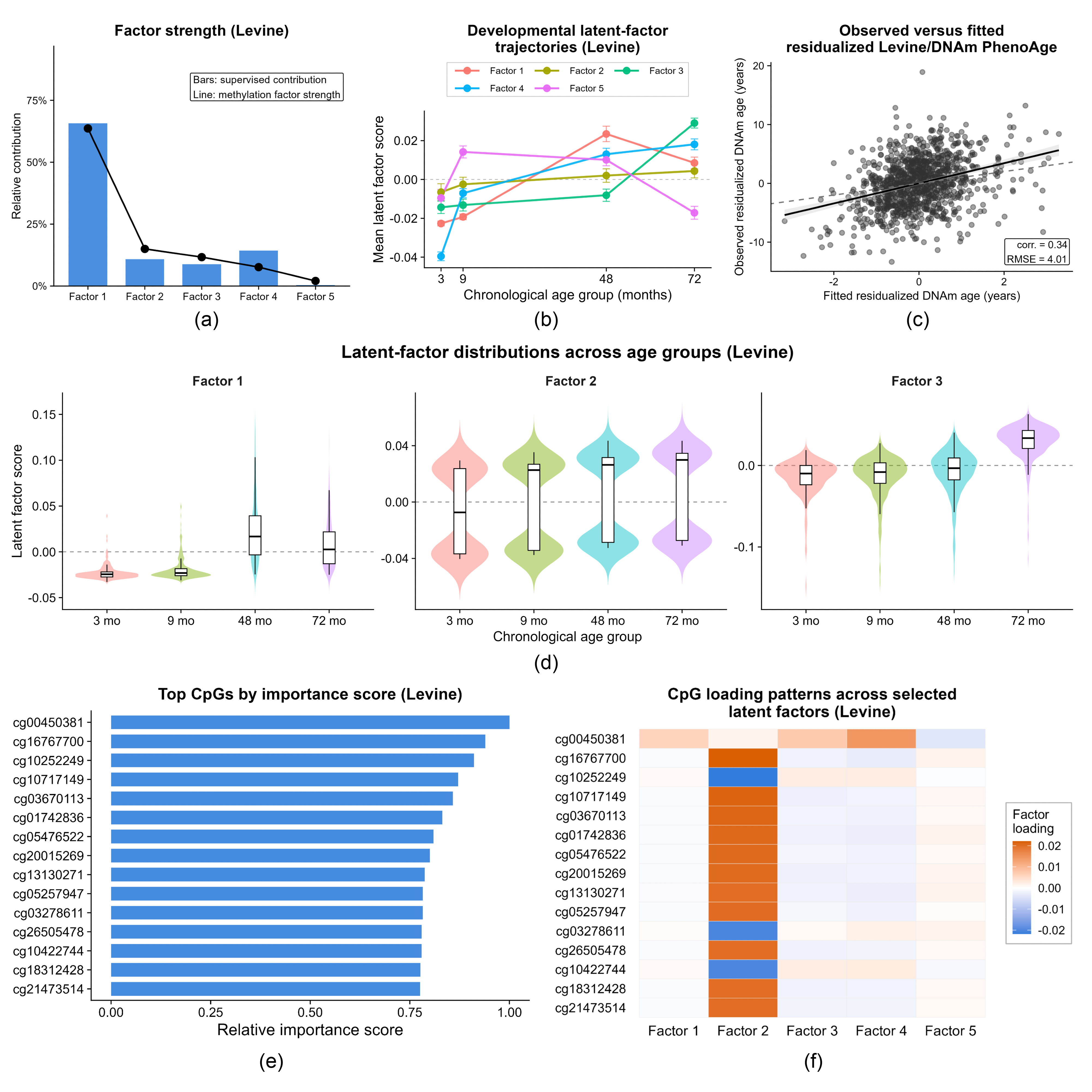}
\caption{
Additional SOLAR analyses for residualized Levine/DNAm PhenoAge. 
(a) Relative supervised contributions of the inferred latent methylation factors together with their corresponding latent-factor strengths derived from the singular-value structure.
(b) Mean developmental trajectories of the inferred latent factors across chronological age groups, with error bars denoting standard errors. 
(c) Observed versus fitted residualized Levine/DNAm PhenoAge values from the fitted SOLAR model. 
(d) Distribution of selected latent-factor scores across developmental age groups, illustrating broader developmental heterogeneity and partially bimodal latent structure relative to the primary Horvath analyses. 
(e) Top 15 CpGs identified by the proposed supervised relative importance score (RIS). 
(f) Loading patterns of the leading CpGs across the inferred latent factors, demonstrating stronger concentration of several CpGs within a dominant latent methylation factor.
}
\label{fig:levine_combined}
\end{figure}

Supplementary Figure~\ref{fig:levine_combined} presents parallel SOLAR analyses using residualized Levine/ DNAm PhenoAge. Relative to the primary Horvath analyses, the residualized Levine/DNAm PhenoAge outcome exhibited broader residual variability, more diffuse developmental latent trajectories, and stronger concentration of several leading CpGs within a dominant latent factor. Nevertheless, the inferred latent structure continued to demonstrate stable low-dimensional organization and partially consistent CpG loading patterns across developmental stages. Supplementary Tables~\ref{tab:Top50CpG_Horvath} and~\ref{tab:Top50CpG_Levine} additionally report the top 50 CpGs identified under  the residualized Horvath DNAm age and residualized Levine/DNAm PhenoAge analyses, respectively, together with their supervised relative importance scores (RIS), dominant latent factors, loading magnitudes, and annotated genes.

To further document the exploratory GO enrichment analysis reported in the main text, Supplementary Table~\ref{tab:solar_go_enrichment} provides the GO biological-process terms corresponding to the leading SOLAR-ranked CpGs from the primary Horvath analysis. The table reports the GO identifier, enrichment ratio, gene ratio, background ratio, adjusted (p)-value, the number of mapped genes and CpGs, and the annotated genes contributing to each term. CpGs were mapped to genes using the \texttt{UCSC\_RefGene\_Name} field from the Illumina EPIC annotation. Because the GO analysis is based on unreplicated exploratory CpG rankings, these results are intended as biological characterization of the inferred supervised methylation structure rather than confirmatory pathway evidence.

\begin{table}[!t]
\centering
\renewcommand{\arraystretch}{1.1}
\resizebox{0.8\columnwidth}{!}{%
\begin{tabular}{|c|c|c|c|c|c|}
\hline
Rank & CpG & \begin{tabular}[c]{@{}c@{}}Relative importance\\ score (RIS)\end{tabular} & Dominant factor & Dominant loading & Annotated gene(s) \\ \hline
1 & cg00450381 & 1.000 & Factor 4 & -0.01443 & EYS \\ \hline
2 & cg20408776 & 0.929 & Factor 1 & -0.0064 & FMNL1 \\ \hline
3 & cg24290948 & 0.927 & Factor 1 & -0.00638 & TMEM40 \\ \hline
4 & cg09268718 & 0.882 & Factor 1 & -0.00625 & SCARF1 \\ \hline
5 & cg19955173 & 0.842 & Factor 1 & -0.0061 & TNK2 \\ \hline
6 & cg13935577 & 0.839 & Factor 1 & -0.00598 & BTBD11 \\ \hline
7 & cg21617353 & 0.817 & Factor 1 & -0.006 & FMNL1 \\ \hline
8 & cg26058474 & 0.814 & Factor 1 & -0.00597 & SNED1 \\ \hline
9 & cg12835012 & 0.808 & Factor 5 & -0.0201 & NA \\ \hline
10 & cg01089914 & 0.803 & Factor 4 & -0.00918 & NA \\ \hline
11 & cg24716664 & 0.793 & Factor 1 & -0.00593 & AP2A2 \\ \hline
12 & cg11209549 & 0.793 & Factor 1 & -0.00584 & TLR5 \\ \hline
13 & cg19452633 & 0.778 & Factor 1 & -0.00587 & HDAC5 \\ \hline
14 & cg03985520 & 0.777 & Factor 1 & -0.00585 & NA \\ \hline
15 & cg09490371 & 0.772 & Factor 1 & -0.00582 & ECEL1P2 \\ \hline
16 & cg13340765 & 0.769 & Factor 1 & 0.00584 & NA \\ \hline
17 & cg01152726 & 0.752 & Factor 1 & -0.00577 & LAMA3 \\ \hline
18 & cg02787560 & 0.749 & Factor 1 & -0.00575 & NA \\ \hline
19 & cg18243764 & 0.746 & Factor 1 & -0.00575 & MANBAL \\ \hline
20 & cg17970299 & 0.738 & Factor 1 & -0.00572 & ZNF385A \\ \hline
21 & cg08457158 & 0.736 & Factor 1 & -0.00571 & NA \\ \hline
22 & cg18454133 & 0.728 & Factor 1 & -0.00568 & EFS \\ \hline
23 & cg18119644 & 0.728 & Factor 1 & -0.00568 & BRSK2 \\ \hline
24 & cg03902220 & 0.727 & Factor 1 & -0.00561 & MAD1L1 \\ \hline
25 & cg04770195 & 0.725 & Factor 1 & -0.00566 & CHD9 \\ \hline
26 & cg10592478 & 0.723 & Factor 1 & -0.00567 & RARG \\ \hline
27 & cg12526693 & 0.723 & Factor 1 & -0.00567 & NA \\ \hline
28 & cg24409870 & 0.714 & Factor 1 & -0.00562 & TSPAN9 \\ \hline
29 & cg07657463 & 0.713 & Factor 1 & -0.00562 & NSUN7 \\ \hline
30 & cg05881698 & 0.711 & Factor 1 & -0.0056 & NA \\ \hline
31 & cg06511312 & 0.710 & Factor 1 & -0.0056 & MEGF6 \\ \hline
32 & cg09598378 & 0.709 & Factor 1 & -0.0056 & XRCC1; PINLYP \\ \hline
33 & cg13475583 & 0.709 & Factor 5 & -0.0058 & NA \\ \hline
34 & cg03339817 & 0.706 & Factor 1 & -0.0056 & SFT2D3 \\ \hline
35 & cg23879118 & 0.705 & Factor 5 & -0.02178 & ACY3 \\ \hline
36 & cg17382258 & 0.703 & Factor 4 & -0.01207 & NA \\ \hline
37 & cg17804342 & 0.703 & Factor 1 & -0.00556 & RGS10 \\ \hline
38 & cg10047182 & 0.695 & Factor 1 & -0.00555 & HLCS \\ \hline
39 & cg21522892 & 0.690 & Factor 1 & -0.00553 & NA \\ \hline
40 & cg13241681 & 0.687 & Factor 1 & -0.00552 & TMCO3 \\ \hline
41 & cg20833492 & 0.687 & Factor 1 & -0.00552 & ZC3H7A \\ \hline
42 & cg04001668 & 0.685 & Factor 1 & -0.00551 & GPR56 \\ \hline
43 & cg08379738 & 0.678 & Factor 1 & -0.00547 & DENND1C \\ \hline
44 & cg21660452 & 0.671 & Factor 1 & -0.00545 & NRXN2 \\ \hline
45 & cg20324784 & 0.670 & Factor 1 & -0.00544 & NA \\ \hline
46 & cg15834833 & 0.669 & Factor 1 & -0.00545 & INHBA-AS1 \\ \hline
47 & cg01030404 & 0.662 & Factor 1 & -0.00542 & NA \\ \hline
48 & cg08101036 & 0.662 & Factor 1 & -0.00542 & HOXA3 \\ \hline
49 & cg02807227 & 0.659 & Factor 1 & -0.00541 & SH3BP4 \\ \hline
50 & cg25834768 & 0.652 & Factor 1 & -0.00537 & CAMKK1 \\ \hline
\end{tabular}}
\caption{Top 50 CpGs ranked by SOLAR relative importance score (RIS) for residualized Horvath DNAm age. Dominant factor denotes the latent factor with the largest absolute loading magnitude for the corresponding CpG.}
\label{tab:Top50CpG_Horvath}
\end{table}

\begin{table}[!t]
\centering
\renewcommand{\arraystretch}{1.1}
\resizebox{0.8\columnwidth}{!}{%
\begin{tabular}{|c|c|c|c|c|c|}
\hline
Rank & CpG & \begin{tabular}[c]{@{}c@{}}Relative importance\\ score (RIS)\end{tabular} & Dominant factor & Dominant loading & Annotated gene(s) \\ \hline
1 & cg00450381 & 1.000 & Factor 4 & 0.01443 & EYS \\ \hline
2 & cg16767700 & 0.939 & Factor 2 & 0.0221 & CNKSR2 \\ \hline
3 & cg10252249 & 0.911 & Factor 2 & -0.02178 & NA \\ \hline
4 & cg10717149 & 0.871 & Factor 2 & 0.0214 & SLC25A14 \\ \hline
5 & cg03670113 & 0.858 & Factor 2 & 0.02123 & TAZ; DNASE1L1 \\ \hline
6 & cg01742836 & 0.832 & Factor 2 & 0.02081 & FHL1 \\ \hline
7 & cg05476522 & 0.810 & Factor 2 & 0.02065 & FGD1 \\ \hline
8 & cg20015269 & 0.800 & Factor 2 & 0.02047 & CLCN5 \\ \hline
9 & cg13130271 & 0.788 & Factor 2 & 0.02024 & PCSK1N \\ \hline
10 & cg05257947 & 0.782 & Factor 2 & 0.02029 & MSL3 \\ \hline
11 & cg03278611 & 0.782 & Factor 2 & -0.02033 & NLGN4Y \\ \hline
12 & cg26505478 & 0.780 & Factor 2 & 0.02024 & CUL4B \\ \hline
13 & cg10422744 & 0.779 & Factor 2 & -0.02007 & NA \\ \hline
14 & cg18312428 & 0.776 & Factor 2 & 0.0202 & MIR718; IRAK1 \\ \hline
15 & cg21473514 & 0.776 & Factor 2 & 0.02023 & CXorf39 \\ \hline
16 & cg23767642 & 0.767 & Factor 2 & 0.02007 & CASK \\ \hline
17 & cg13574945 & 0.767 & Factor 2 & 0.02003 & OTUD5 \\ \hline
18 & cg08059778 & 0.758 & Factor 2 & 0.01992 & BCORL1 \\ \hline
19 & cg15258447 & 0.757 & Factor 4 & 0.01647 & NA \\ \hline
20 & cg23696472 & 0.757 & Factor 2 & 0.0199 & TSPYL2 \\ \hline
21 & cg05806018 & 0.756 & Factor 2 & 0.01988 & AFF2 \\ \hline
22 & cg02869694 & 0.755 & Factor 2 & 0.01993 & IKBKG; G6PD \\ \hline
23 & cg15925199 & 0.751 & Factor 2 & 0.01973 & OTUD5 \\ \hline
24 & cg27277239 & 0.745 & Factor 2 & 0.01966 & OTUD5 \\ \hline
25 & cg00723973 & 0.745 & Factor 2 & 0.01971 & RP2 \\ \hline
26 & cg09229960 & 0.744 & Factor 2 & 0.01977 & EMD \\ \hline
27 & cg08969352 & 0.727 & Factor 2 & 0.01953 & FANCB; MOSPD2 \\ \hline
28 & cg07837161 & 0.727 & Factor 2 & 0.01961 & FAM3A \\ \hline
29 & cg14248084 & 0.727 & Factor 2 & -0.01954 & NA \\ \hline
30 & cg06136002 & 0.724 & Factor 2 & 0.01948 & HMGB3 \\ \hline
31 & cg22028367 & 0.718 & Factor 2 & -0.01947 & \begin{tabular}[c]{@{}c@{}}FAM197Y2; FAM197Y5;\\  TSPY4\end{tabular} \\ \hline
32 & cg23712855 & 0.716 & Factor 2 & 0.01939 & DKC1 \\ \hline
33 & cg12935118 & 0.708 & Factor 2 & 0.01928 & SMARCA1 \\ \hline
34 & cg02794321 & 0.703 & Factor 2 & 0.01921 & AMMECR1 \\ \hline
35 & cg03391801 & 0.703 & Factor 2 & 0.0192 & EMD \\ \hline
36 & cg07861180 & 0.702 & Factor 2 & 0.01916 & NA \\ \hline
37 & cg17382258 & 0.701 & Factor 4 & 0.01207 & NA \\ \hline
38 & cg20455959 & 0.700 & Factor 2 & 0.01916 & MTMR1 \\ \hline
39 & cg03834574 & 0.696 & Factor 2 & 0.01908 & KLF8 \\ \hline
40 & cg10818284 & 0.696 & Factor 2 & 0.01902 & SYP \\ \hline
41 & cg12124890 & 0.690 & Factor 2 & 0.01908 & CASK \\ \hline
42 & cg24186901 & 0.688 & Factor 2 & 0.01895 & EDA \\ \hline
43 & cg21290550 & 0.683 & Factor 2 & 0.01895 & \begin{tabular}[c]{@{}c@{}}LOC100133957; \\ UXT\end{tabular} \\ \hline
44 & cg05053121 & 0.681 & Factor 2 & -0.01887 & TMSB4Y \\ \hline
45 & cg11846372 & 0.680 & Factor 2 & -0.01886 & TXLNGY \\ \hline
46 & cg23680829 & 0.678 & Factor 2 & 0.01896 & IKBKG; G6PD \\ \hline
47 & cg13023833 & 0.678 & Factor 2 & 0.01883 & ELK1 \\ \hline
48 & cg24627956 & 0.677 & Factor 2 & 0.01884 & NA \\ \hline
49 & cg02195366 & 0.677 & Factor 2 & 0.01885 & MSL3 \\ \hline
50 & cg25205946 & 0.676 & Factor 2 & 0.01885 & WDR45 \\ \hline
\end{tabular}}
\caption{Top 50 CpGs ranked by SOLAR relative importance score (RIS) for residualized Levine/DNAm PhenoAge. Dominant factor denotes the latent factor with the largest absolute loading magnitude for the corresponding CpG.}
\label{tab:Top50CpG_Levine}
\end{table}

\begin{sidewaystable}[!p]
\centering
\scriptsize
\setlength{\tabcolsep}{2.5pt}
\renewcommand{\arraystretch}{0.99}

\begin{adjustbox}{max width=\textheight, max totalheight=\textwidth, keepaspectratio}
\begin{tabular}{p{1.8cm} p{2.2cm} p{1.1cm} p{1.8cm} p{1.9cm} p{1.0cm} p{1.4cm} p{1.4cm} p{1.4cm} p{1.2cm} p{1.2cm} p{4.8cm}}
\toprule
GO ID & GO term & Gene Ratio & BgRatio & Enrichment ratio & Count & $p$-value & Adjusted $p$-value & $q$-value & Unique genes & Unique CpGs & Gene names \\
\midrule
GO:0007015 & actin filament organization & 32/501 & 448/18842 & 2.686 & 32 & $4.45e{-7}$ & 0.00262 & 0.00262 & 32 & 41 & ARHGAP25; CCL24; CORO2B; CORO7; CX3CL1; DNM2; EFHD2; ELMO2; ELMO3; EPHA1; EPS8; F11R; FLII; FMN1; GAS2; HCK; LIMA1; LIMK1; MICALL2; MYO5C; NF2; RAC1; SH3BP4; SHROOM3; SMAD3; SRC; SVIL; SWAP70; SYNPO; TGFBR1; TMOD3; TPM4 \\

GO:0032233 & positive regulation of actin filament bundle assembly & 10/501 & 61/18842 & 6.165 & 10 & $4.31e{-6}$ & 0.00591 & 0.00591 & 10 & 12 & CX3CL1; EPHA1; LIMA1; LIMK1; NF2; RAC1; SMAD3; SWAP70; SYNPO; TGFBR1 \\

GO:0043297 & apical junction assembly & 10/501 & 62/18842 & 6.066 & 10 & $5.02e{-6}$ & 0.00591 & 0.00591 & 10 & 14 & CTNNA1; ECT2; F11R; MICALL2; PECAM1; PKN2; PRKCH; RAC1; TBCD; VCL \\

GO:0051552 & flavone metabolic process & 5/501 & 10/18842 & 18.804 & 5 & $2.94e{-6}$ & 0.00591 & 0.00591 & 5 & 5 & UGT1A10; UGT1A6; UGT1A7; UGT1A8; UGT1A9 \\

GO:0071384 & cellular response to corticosteroid stimulus & 10/501 & 60/18842 & 6.268 & 10 & $3.69e{-6}$ & 0.00591 & 0.00591 & 10 & 11 & HCN2; SCNN1A; SSTR5; UBE2L3; UGT1A10; UGT1A6; UGT1A7; UGT1A8; UGT1A9; ZFP36L1 \\

GO:0014009 & glial cell proliferation & 8/501 & 41/18842 & 7.338 & 8 & $1.04e{-5}$ & 0.00894 & 0.00894 & 8 & 8 & CHRM1; CX3CL1; E2F1; IL34; NF2; PRKCH; PRKCI; SKI \\

GO:0031589 & cell-substrate adhesion & 24/501 & 333/18842 & 2.711 & 24 & $1.06e{-5}$ & 0.00894 & 0.00894 & 24 & 30 & AJAP1; CDK6; CORO2B; CX3CL1; DDR1; DOCK5; EPHA1; ITGA9; ITGB2; LAMB1; LAMB3; MICALL2; MSLN; MSLNL; NF2; PKD1; RAC1; SKAP1; SMAD3; SNED1; SRC; TBCD; TIAM1; VCL \\

GO:0007160 & cell-matrix adhesion & 19/501 & 229/18842 & 3.120 & 19 & $1.26e{-5}$ & 0.00928 & 0.00928 & 19 & 22 & AJAP1; CDK6; CORO2B; CX3CL1; DDR1; EPHA1; ITGA9; ITGB2; MSLN; MSLNL; NF2; PKD1; RAC1; SKAP1; SMAD3; SNED1; SRC; TIAM1; VCL \\

GO:0007264 & small GTPase-mediated signal transduction & 30/501 & 490/18842 & 2.303 & 30 & $2.17e{-5}$ & 0.01150 & 0.01150 & 30 & 36 & ARHGAP25; ARHGEF16; ARHGEF3; DOCK5; DOCK9; ECT2; EPS8; ERBB2; F11R; GARNL3; IQSEC3; LIMK1; MICALL2; NISCH; OBSCN; PKP4; PSD3; RAB35; RAC1; RALGPS1; RASA3; RASGRP2; RIT1; SH2D3C; SH3BP4; SRC; STARD13; SWAP70; TIAM1; TNK2 \\

GO:0014855 & striated muscle cell proliferation & 9/501 & 58/18842 & 5.836 & 9 & $2.07e{-5}$ & 0.01150 & 0.01150 & 9 & 9 & FGFR2; NRG1; PPARD; RUNX1; SKI; STAT3; TGFBR1; TP73; YAP1 \\

\bottomrule
\end{tabular}
\end{adjustbox}
\caption{GO biological-process enrichment analysis of the top 1000 SOLAR-ranked CpGs. CpGs were mapped to genes using the \texttt{UCSC\_RefGene\_Name} annotation field from the Illumina EPIC annotation. GeneRatio denotes the number of genes in the GO term divided by the number of unique mapped input genes used in the enrichment analysis.}
\label{tab:solar_go_enrichment}
\end{sidewaystable}

\clearpage
\bibliographystyle{agsm}
\bibliography{MM-MC}